\documentclass[journal]{IEEEtran}

\usepackage[hyphens]{url}
\usepackage[hidelinks]{hyperref}
\usepackage{float}

\usepackage{comment}
\usepackage{graphicx}
\usepackage{listings}

\usepackage[inline]{enumitem}
 
\usepackage{listings}
\usepackage{xcolor}

\definecolor{eclipseStrings}{RGB}{42,0.0,255}
\definecolor{eclipseKeywords}{RGB}{127,0,85}
\colorlet{numb}{magenta!60!black}

\lstdefinelanguage{json}{
    basicstyle=\normalfont\ttfamily,
    commentstyle=\color{eclipseStrings}, %
    stringstyle=\color{eclipseKeywords}, %
    numbers=left,
    numberstyle=\scriptsize,
    stepnumber=1,
    numbersep=8pt,
    showstringspaces=false,
    breaklines=true,
    frame=lines,
    backgroundcolor=\color{white}, %
    string=[s]{"}{"},
    comment=[l]{:\ "},
    morecomment=[l]{:"},
    literate=
        *{0}{{{\color{numb}0}}}{1}
         {1}{{{\color{numb}1}}}{1}
         {2}{{{\color{numb}2}}}{1}
         {3}{{{\color{numb}3}}}{1}
         {4}{{{\color{numb}4}}}{1}
         {5}{{{\color{numb}5}}}{1}
         {6}{{{\color{numb}6}}}{1}
         {7}{{{\color{numb}7}}}{1}
         {8}{{{\color{numb}8}}}{1}
         {9}{{{\color{numb}9}}}{1}
}
\usepackage{amsmath}

\usepackage{booktabs} %

\usepackage[linesnumbered,ruled]{algorithm2e}
\SetKwRepeat{Do}{do}{while}

\SetCommentSty{mycommfont}

\usepackage{
authblk}

\usepackage{tabularx} %
\usepackage{ragged2e}
\usepackage{booktabs} %

\usepackage{cite}
\newcommand{\component}[1]{\textsl{#1}}

\newcommand{\region}[1]{\texttt{#1}}
\newcommand{\csp}[1]{\texttt{#1}}

\newcommand{\RG}[1]{\texttt{#1}}

\newcounter{example}[section]

\begin{document}
\bstctlcite{IEEEexample:BSTcontrol}

\title
{A Reference Architecture for Governance of Cloud Native Applications}
\author[1]{William Pourmajidi}
\author[2]{Lei Zhang}
\author[3]{John Steinbacher}
\author[4]{Tony Erwin}
\author[1]{Andriy Miranskyy}
\affil[1]{Toronto Metropolitan University, Toronto, Canada}
\affil[2]{University of Maryland, Baltimore County, USA}
\affil[3]{IBM Canada Lab, Toronto, Canada}
\affil[4]{IBM Cloud Platform, Austin, USA}
\affil[ ]{william.pourmajidi@torontomu.ca, leizhang@umbc.edu, jstein@ca.ibm.com, aerwin@us.ibm.com, avm@torontomu.ca }

\maketitle
\begin{abstract}
The evolution of cloud computing has given rise to Cloud Native Applications (CNAs), presenting new challenges in governance, particularly when faced with strict compliance requirements. This work explores the unique characteristics of CNAs and their impact on governance. We introduce a comprehensive reference architecture designed to streamline governance across CNAs, along with a sample implementation, offering insights for both single and multi-cloud environments. Our architecture seamlessly integrates governance within the CNA framework, adhering to a ``battery-included'' philosophy. Tailored for both expansive and compact CNA deployments across various industries, this design enables cloud practitioners to prioritize product development by alleviating the complexities associated with governance. In addition, it provides a building block for academic exploration of generic CNA frameworks, highlighting their relevance in the evolving cloud computing landscape.

\end{abstract}


\IEEEpeerreviewmaketitle

\section{Introduction}\label{sec:Introduction}

Cloud computing has become a primary platform for businesses of all sizes. As predicted by the International Data Corporation FutureScape, 80\% of enterprises will put a mechanism in place to shift to cloud-centric infrastructure and applications twice as fast as before the pandemic~\cite{IDCFutureScape:online}. cloud's key features, such as scalability, elasticity, ubiquitous access, and pay-as-you-go, coupled with reduced operational costs, render cloud computing appealing to a wide array of organizations~\cite{hsu2014examining,nandgaonkar2014comprehensive}.

The National Institute of Standards and Technology (NIST) interprets cloud governance as a standardization focus, akin to Internet governance, aiming to ensure interoperability and data portability across various cloud platforms~\cite{NISTClou61:online}. This perspective anticipates future standards for efficient data and service access tailored to CSCs' needs~\cite{hogan2011nist}. Red Hat describes cloud governance as the framework of policies guiding an organization's cloud operations~\cite{redhat-cloudgovdefinition}. AWS defines cloud governance as a system allowing customers to set and monitor their cloud requirements in terms of security, cost, and oversight~\cite{AWSCloud85:online}. Azure views governance as a tool for safe cloud adoption, balancing control, security, and compliance with cloud scalability~\cite{Governme20:online}. Google Cloud specializes in its governance definition of data management, ensuring data security, privacy, and compliance with internal and external standards~\cite{WhatisDa76:online}. IBM considers governance as practices to balance benefits, risk, and resources in cloud environments~\cite{IBMCloud51:online, Whatisda59:online}. In this work, \textbf{governance} encompasses the management of all aspects of cloud computing and applications, including CNAs, covering functions like monitoring, observability, and compliance for comprehensive cloud management.

\subsection{Cloud Computing Evolution}\label{subsec:Cloud Computing Evolution}
The past two decades have seen numerous enhancements in cloud services, underpinned by shifts in practices among CSPs and CSCs. These enhancements are largely attributable to changes in cloud infrastructure and cloud application architecture~\cite{varghese2018next}. 

 \subsubsection{Cloud Infrastructure}

The underlying cloud computing infrastructure and technologies surrounding it are constantly changing. The so-called software-defined cloud computing is becoming increasingly popular, and CSPs use advanced virtualization techniques to convert all data center resources to virtualized ones. For example, software-defined networks are replacing traditional networks, and software-defined storage solutions are replacing traditional storage options. On a macro-level, CSPs are evolving their infrastructure management approaches to achieve software-defined data centers~\cite{abbasi2019software}. 

\subsubsection{Cloud Application Architecture}

Cloud application architectures have transformed from monolithic to dynamic, microservice-based designs, meeting cloud environments' scalability and flexibility needs~\cite{WhatisaC94:online}. This shift necessitates lightweight and efficient computing, making containers the preferred choice for their microservice compatibility~\cite{liu2020microservices}. Consequently, Kubernetes has become essential for managing and orchestrating microservices~\cite{Kubernet86:online}. Moreover, the rise of Serverless computing offers an event-driven model, enhancing efficiency by utilizing resources only as needed, in contrast to traditional Virtual Machines (VMs), which often require continuous resource allocation regardless of usage, impacting financial feasibility and resource utilization~\cite{van2018serverless}.

\subsection{The Rise of CNAs}\label{subsec:The Rise of CNAs}

CNAs have emerged as a modern class of cloud applications, inherently designed to leverage the unique capabilities of cloud environments. The Cloud Native Computing Foundation (CNCF) defines CNAs as software systems specifically built to exploit cloud innovations, emphasizing their potential to deliver resiliency and scalability when designed accordingly~\cite{CloudNat91:online}. Given their increasing importance, CNAs have become a critical focus area in cloud computing. However, the lack of comprehensive governance architectures presents significant risks in their design and implementation~\cite{odun2019systematic,alonso2019decide,alonso2023understanding}. This paper introduces a reference architecture to address these governance challenges. For a more detailed discussion of CNA characteristics, please refer to Section~\ref{subsec:Characteristics of CNAs}.

\subsection{CNA Governance}\label{subsec:CNA Governance}

CNA Governance, drawing inspiration from cloud and IT governance, involves strategic, architectural, operational, and compliance elements within CNAs~\cite{9284234}. It aligns CNAs' use with business goals and regulatory requirements, incorporating necessary policies and guidelines for resource management. Governance also encompasses security for threat protection, compliance for legal adherence, and performance management for meeting Service Level Agreements (SLAs). Particularly in regulated industries, a CNA cannot operate unless its adherence to all aspects of compliance requirements is met, making governance an indispensable aspect.

CNAs bring complex governance challenges due to their microservice architectures and deployment in orchestrated environments like Kubernetes clusters~\cite{Kubernet86:online}. They offer agility and scalability, key attributes augmented by serverless computing. While serverless computing efficiently optimizes resource usage, it also introduces additional complexities in governing ephemeral resources~\cite{van2018serverless}. Another layer of complexity in CNA governance stems from the need to implement distinct scalability strategies for both stateless and stateful components.

Moreover, the continuous evolution of CNAs is driven by methodologies such as Continuous Integration and Continuous Delivery (CI/CD) and practices like Everything as Code (XaC) and Infrastructure as Code (IaC). Tools like Terraform~\cite{Terrafor55:online}, Chef~\cite{ChefSoft16:online}, and Ansible~\cite{Infrastr7:online} are integral to these methodologies, facilitating consistent and automated deployment. This ever-changing landscape of CNAs underscores the need for an adaptable and continuous governance framework. Our proposed reference architecture is designed to meet these dynamic requirements, providing a comprehensive solution to the governance challenges of CNAs. For readers seeking a more in-depth exploration of CNA governance challenges, please refer to Section~\ref{subsec:CNA Governance Challenges} for further details.

\subsection{Our Contribution}
\label{subsec:Our Contribution}
This paper navigates the complexities of governing CNAs by developing a targeted reference architecture for CNA governance. The proposed reference architecture and its reference implementation on AWS and IBM public clouds are designed to simplify governance processes in CNAs, particularly critical for regulated industries. By embedding essential governance aspects like monitoring, observability, and compliance into the architecture, we significantly reduce the complexity for cloud practitioners when integrating these elements into their CNAs. This integration fosters a comprehensive and efficient governance model tailored to the unique demands of CNAs.

Our investigation is guided by three Research Questions~(RQs) as follows.

\begin{itemize}
\item \textbf{RQ1}. What distinct characteristics define CNAs?

\item \textbf{RQ2}. What are the desired characteristics of a reference architecture for CNA governance? 

\item \textbf{RQ3}. How can the software architecture guide the creation of a custom reference architecture for governing CNAs, integrating RQ1 characteristics and RQ2 components?
\end{itemize}

To address these research questions, we employed a methodology grounded in software engineering principles, specifically through the lens of functional and non-functional requirements analysis. Our approach involved systematically analyzing the challenges presented by each RQ and iteratively designing a governance-oriented reference architecture. This methodology enabled us to balance theoretical insights with practical design considerations while ensuring that the resulting architecture remains extensible, implementation-agnostic, and suitable for multi-cloud environments.

The remainder of this paper is organized as follows. In Section~\ref{sec:Background and Motivation}, we dive into the characteristics of CNAs, addressing RQ1 by elaborating on what precisely defines these applications. This section also explores the challenging landscape of CNA governance, offering insights into our first research question. We continue in Section~\ref{sec:Desired Characteristics of a Reference Architecture for CNA Governance} by discussing the desired characteristics of a reference architecture for the governance of CNAs, thus answering RQ2. This exploration leads to Section~\ref{sec:Proposed Reference Architecture}, where we present our proposed reference architecture in detail, answering RQ3 and using the insights from the preceding sections. To demonstrate practical applicability, 
Section~\ref{sec:Real-world Instantiation of the Proposed Reference Architecture} provides real-world examples of how this architecture could be instantiated. A review of related literature, covering CNAs and their governance challenges, is presented in Section~\ref{sec:Related Literature}. We bring the paper to a close in Section~\ref{sec:Conclusion}, summarizing our findings and contributions.

\section{Background and Motivation}\label{sec:Background and Motivation}

This section serves as the foundation for our exploration, delving into the defining characteristics of CNAs and the inherent governance challenges they present. In Section~\ref{subsec:Characteristics of CNAs}, we identify key aspects that distinguish CNAs, addressing our RQ1 and laying the groundwork for understanding their governance complexities. Following this, Section~\ref{subsec:CNA Governance Challenges} examines the challenges in governing these dynamic and scalable applications, while Section~\ref{subsec:CNA Governance Aspects} explores broader cloud governance aspects. Together, these insights form the motivation for our work, explaining why a tailored governance approach for CNAs is essential.

\subsection{Characteristics of CNAs}\label{subsec:Characteristics of CNAs}

CNAs are exclusively designed, developed, deployed, and maintained for cloud environments, fully utilizing cloud-specific resources and services. Contrary to common assumptions, merely running applications on virtualized cloud infrastructure does not qualify them as CNAs. Virtualization and cloud infrastructure are essential but not exhaustive components of CNAs. Distinguishing between generic cloud-based applications and true CNAs is vital for implementing effective governance solutions tailored to the nature of the deployed application~\cite{leymann2016native}. In this section, we outline and discuss the defining characteristics of CNAs to aid practitioners in this differentiation.

\subsubsection{Microservice architecture} The defining attribute of CNAs lies in their microservice architecture~\cite{kratzke2017understanding,balalaie2015migrating}. This architectural style involves decomposing large, complex applications into atomic, isolated services. These services, each responsible for a specific piece of business logic, communicate with each other predominantly through lightweight, often RESTful, APIs~\cite{Microser77:online}. This encourages a modular approach, where components can be developed, deployed, and scaled independently. Such decentralization not only increases agility but is also crucial for scaling applications efficiently within CNAs~\cite{zhang2019microservice}. The microservice architecture offers significant benefits in terms of flexibility and scalability, aligning with the dynamic nature of cloud environments and the continuous integration and delivery models they often employ.

\subsubsection{Containers as computing platform} In line with the microservice architecture of CNAs, containers emerge as a fitting computing platform. Given that each microservice component is responsible for atomic business logic and requires limited computational power, containers offer a more appropriate solution than VMs~\cite{singh2017container,amaral2015performance}. This suitability stems from containers' ability to be easily managed, replicated, and scaled, in contrast to VMs, which might provide more resources than necessary for microservices. The shared core operating system kernel in containers not only facilitates efficient hardware use but also aligns with the principles of cloud computing that prioritize optimal resource utilization and efficiency. Recognizing this, CSPs have advanced their services, including AWS Elastic Kubernetes Service~\cite{ManagedK42:online}, Azure Kubernetes Services~\cite{ManagedK46:online}, Google cloud Kubernetes Engine~\cite{Kubernet19:online}, and IBM Cloud Kubernetes Service~\cite{IBMCloud51:online}, to provide CSCs with robust options for managing and orchestrating containers, thereby streamlining the implementation of microservices in CNAs.

\subsubsection{Serverless as computing platform} CNAs are composed of various components with different run-time requirements. For instance, while the user-interface components and Online Transaction Processing services need to be up and running almost $24/7$, batch processing tasks, aggregations, and backup operations demand a shorter and less frequent run-time. Thus, dedicating a VM or container to a service that runs once or a few times a day is an infeasible under-utilization~\cite{li2009method}. The better utilization of resources on the CSPs translates to a more feasible solution for the CSCs~\cite{eivy2017wary}. CSPs have brought up a new computing platform known as serverless to address schedulable workloads. Serverless computing hides server usage from developers and runs the associated code on-demand, only charging the user based on the time that code is running~\cite{castro2019rise}. Interestingly, serverless computing and microservices have a lot of common characteristics. If a microservice does not need to be up and running at all times, a serverless computing platform can bring even more financial feasibility than containers~\cite{taibi2020serverless, WhatareC56:online}. CSPs have developed and released services such as AWS Lambda~\cite{Serverle63:online}, Azure Serverless~\cite{AzureSer62:online}, Google cloud Serverless Computing~\cite{Serverle17:online}, and IBM Cloud Code Engine~\cite{IBMCloud93:online}.

\subsubsection{State isolation} CNAs are built and implemented with a clear separation between stateless and stateful services. In the context of CNAs, even if stateless services interact with stateful services, this interaction is often managed via a set of exposed APIs~\cite{wurster2020cloud}. This separation allows different scalability practices for each stateless and stateful component. Stateless components, often implemented as containers or serverless functions, require less effort for scalability and reliability. In contrast, stateful component scalability and reliability are not as straightforward and require architectural consideration, including replication and storing the state at any given time. An isolated state is one of the desired characteristics of the CNAs~\cite{kratzke2017understanding, fehling2014cloud}.

\subsection{CNA Governance Challenges}\label{subsec:CNA Governance Challenges}

In addressing CNA governance challenges, we encounter complexities across architectural components and data types. This includes microservice distribution, container dynamics, serverless platform limitations, diverse telemetry and their Big Data trail, and infrastructure governance intricacies. Each element poses unique governance hurdles, and we briefly explore these to understand effective CNA governance.

\subsubsection{Microservice challenges} Microservice architecture imposes a few different types of governance challenges. Given their distributed nature, each microservice is deployed on an isolated computing platform such as a container. Often, there is a high degree of dependency among microservices, and failure in one may cause a ripple effect, and the entire system may be impacted. Hence, in addition to the internal governance of the microservices and their computing platform, one needs to consider the complex relationships among them to construct a representative view of the entire deployed CNA~\cite{levin2020viperprobe}.

\subsubsection{Container challenges} Containers are among the essential building blocks of CNAs. Microservices are often deployed on a wide range of containers that may or may not be part of the same cluster. Collecting and analyzing telemetry from widely distributed containers is already a challenging task~\cite{moradi2017conmon}. As the number and nature of the containers running a particular CNA are dynamic, the governance solution must be aware of any potential changes to the infrastructure~\cite{ raj2018automated}. That is why many practitioners use governance tools such as Kubernetes on top of container orchestration solutions~\cite{martin2021observability}.

\subsubsection{Serverless challenges} CSPs that offer Function-as-a-Service (FaaS), also known as serverless platforms, often provide generic dashboards to monitor the current state of processes running on FaaS platforms. However, these dashboards offer limited features such as visualizing the workload, reports on resource utilization, and run-time logs. As the root access to the underlying operating system is not provided in a FaaS offering, installation and configuration of any additional governance tool on them is not possible~\cite{cordingly2021enhancing}.

\subsubsection{Diverse CNA's telemetry} The array of computing platforms and cloud services in CNAs results in a wide range of telemetry data, including logs, metrics, and traces. Processing this heterogeneous data for meaningful analysis involves cleaning, formatting, and aligning it using time as a unifying element. This cross-referencing is crucial for effective observability in CNAs, facilitating the management of their operational health. Addressing the complexity of diverse telemetry data is essential, as highlighted in several studies, underscoring the need for methodical and standardized data management~\cite{gatev2021observability, miranskyy2007iterative, miranskyy2008sift, MetricsL15:online}.

\subsubsection{Big Data characteristics of CNA's telemetry} The extensive scale of cloud components in CNAs generates a vast volume of data, exemplifying the 5Vs of Big Data---Volume, Velocity, Variety, Veracity, and Value. This governance data, featuring high volume and velocity, is derived from a wide variety of sources, each contributing to its complexity. The veracity of this data is vital, as it must be accurate and reliable for effective decision-making. The value derived from this telemetry data is pivotal for enhancing the governance and operational efficiency of CNAs. This substantial data stream, marked by its Big Data attributes, necessitates robust solutions for storage, processing, and analysis to effectively manage and utilize this wealth of information for governance purposes. The challenges in handling such Big Data aspects of cloud telemetry have been explored in existing research, highlighting the need for sophisticated and scalable solutions tailored to the unique requirements of CNAs~\cite{miranskyy2016operational, pourmajidi2017challenges, islam2020anomaly, islam2021anomaly, pourmajidi2019dogfooding, pourmajidi2021challenging, sohana2024cloudheatmap, saiful2024anomaly}.

\subsubsection{Infrastructure governance challenges} Deploying CNAs effectively on Infrastructure as a Service (IaaS) and Platform as a Service (PaaS) demands thorough infrastructure governance, which is central to comprehensive cloud governance. In these environments, practitioners must incorporate governance solutions, key for monitoring and managing CNA deployments. This governance aspect entails synthesizing diverse telemetry data from numerous sources to create an accurate representation of the CNA's operational state. The complexity of managing CNA infrastructure, given the variety of data sources and the depth of information, is critical. Previous research has explored these challenges, underlining the evolving demands of infrastructure monitoring in the context of CNAs~\cite{pourmajidi2017challenges, pourmajidi2021challenging, pourmajidi2019dogfooding}.

\subsection{CNA Governance Aspects}\label{subsec:CNA Governance Aspects}

In this section, we discuss key aspects of CNA governance, highlighting the unique challenges of monitoring, observability, and compliance. These crucial components span a broad spectrum of technical, financial, and legal considerations, forming the foundation of our comprehensive approach to CNA governance. We will explore each area to unravel the complex nature of managing and maintaining CNAs effectively.

\subsubsection{Monitoring challenges} In the dynamic field of cloud computing and cloud governance, CSPs are tasked with ensuring the reliability, scalability, and security of their platforms. Concurrently, CSCs face the challenge of rigorous monitoring to uphold SLAs. This dual responsibility has given rise to sophisticated monitoring needs intertwined with the concept of observability. Observability, in this context, is the capability to infer the status of complex systems through their output data. It is a critical component in the governance of CNAs~\cite{WhatIsCl93:online, esposito2016challenges, yu2016cloudseer, Summaryo12:online, AmazonAn9:online, wahab-aiops}.

\subsubsection{Observability challenges} Observability, a concept foundational to cloud computing governance, plays a vital role in the effective visualization and prediction of system behaviors. It encompasses the collection and analysis of system data, predominantly focusing on logs, metrics, and traces. Crucial to observability is the balance between data comprehensiveness and maintaining optimal system performance. This often involves implementing adaptive sampling techniques, ensuring that data collection is as efficient as it is informative, and facilitating deeper insights into CNA operations~\cite{bohner2012controllability, kalman1960general, Observability}.

\subsubsection{Compliance challenges} In the cloud environment, particularly for CNAs operating in regulated industries, compliance is an essential concern. Solutions deployed must adhere to specific functional and non-functional requirements, often intertwined with SLAs. Observability, therefore, emerges as a key element in ensuring adherence to these stringent requirements. By leveraging observability tools, organizations can continuously monitor and verify their compliance with relevant standards and regulations, ensuring that their cloud solutions remain within the bounds of mandated guidelines~\cite{joshi2020integrated}. Compliance often involves demonstrating that the system meets specific controls outlined in established frameworks such as NIST, FEDRAMP, HIPAA, SOC2, and C5. These frameworks enumerate hundreds of controls across categories like access management, data protection, and incident response. For example, adherence to HIPAA mandates safeguards for patient data, while SOC2 emphasizes security, availability, and confidentiality controls. Observability plays a critical role in continuously verifying that these controls are met, providing organizations with the evidence required to demonstrate compliance under audit and maintain adherence to SLAs. CSPs often employ observability tools to collect evidence and ensure compliance with these frameworks. For instance, IBM utilizes its Security and Compliance Center (SCC) and internal tools like "fetcher" and "checker" to automate evidence collection and verification. These tools exemplify how CSPs can enhance compliance by integrating observability with automated compliance checks, enabling organizations to continuously monitor and demonstrate that required controls are implemented and adhered to effectively.

\section{Desired Characteristics of a Reference Architecture for CNA Governance}\label{sec:Desired Characteristics of a Reference Architecture for CNA Governance}

This section addresses RQ2 by outlining the desired characteristics of a reference architecture tailored to CNA governance, leveraging insights from our extensive experience and empirical research. It includes non-functional requirements in Section~\ref{subsec:Desired Non-Functional Requirements}, architectural requirements in Section~\ref{subsec:Desired Architectural Requirements}, and delves into the construction, deployment, and maintenance requirements in Section~\ref{subsec:Desired Construction, Deployment, and Maintenance Requirements}. Furthermore, key components crucial for effective governance are detailed in Section~\ref{subsec:Desired Components for Governance of CNAs}. This structured approach ensures the architecture supports CNAs optimally, emphasizing performance, reliability, and compliance across various governance facets.

\subsection{Desired Non-Functional Requirements}\label{subsec:Desired Non-Functional Requirements}

For the effective governance of CNAs, a reference architecture must incorporate certain non-functional requirements. Drawing on our extensive experience in CNAs design and implementation, and based on a set of non-functional requirements in software engineering~\cite{NFR-chung2012non, NFR-paradkar2017, NFR-adams2015}, we identify the following set of non-functional requirements as foundational for ensuring CNAs' optimal operation. This list provides a solid foundation but can be adjusted by practitioners to include or exclude additional non-functional requirements as deemed appropriate, ensuring alignment with the specific governance needs and unique characteristics of CNAs.

\subsubsection{Performance and reliability} A desired aspect of the reference architecture is incorporating software-defined load-balancers to evenly distribute workload among clusters. This approach is fundamental to achieving the desired levels of performance and reliability, ensuring CNAs function seamlessly under various operational demands.

\subsubsection{Adaptability} The reference architecture should be inherently adaptable to a range of cloud platforms and CNA configurations. This flexibility allows for modifications to the architecture to address the unique requirements of different deployment environments.

\subsubsection{Extensibility} Extensibility is crucial for a reference architecture to accommodate evolving governance needs. The architecture must be designed with the capacity to be extended and customized, enabling practitioners to tailor it to their specific operational requirements.

\subsubsection{Elasticity} Auto-scaling and Intelligent Workload Management (IWM) are essential for efficient resource utilization. The reference architecture should advocate for elasticity, utilizing auto-scaling and IWM solutions to enforce optimized workload strategies~\cite{nicoletti2018cross} from prominent CSPs such as AWS Auto Scaling~\cite{AWSAutoS97:online}, Azure Autoscale~\cite{AzureAut72:online}, Google cloud Autoscaling~\cite{Autoscal78:online}, and IBM Cloud Autoscaling~\cite{Autoscal15:online} to ensure resources are optimally managed. 

\subsubsection{Integrability} The ability to seamlessly integrate with existing or future platforms is a desired characteristic~\cite{henttonen2007integrability}. The reference architecture needs to ensure components are sufficiently flexible and replaceable, facilitating easy integration and thus enhancing the governance framework's integrability.

\subsubsection{High Availability} In cloud and distributed systems, ensuring high availability is critical and stands as a fundamental non-functional requirement~\cite{mesbahi2018reliability}. A reference architecture must integrate strategies for high availability, including the use of load balancers, clusters, and especially an auxiliary data bus to maintain continuous communication among components, even in the event of primary channel disruptions. Such a design, emphasizing failovers and redundancy upfront, is essential for achieving uninterrupted governance processes and resilience~\cite{sharma2014framework}. This approach underscores the pivotal role of high availability in supporting the robustness and reliability of CNAs.

\subsection{Desired Architectural Requirements}\label{subsec:Desired Architectural Requirements}

For the effective governance of CNAs, the architectural foundation of a reference architecture is critical, including a series of architectural characteristics essential for ensuring optimal flexibility and reliability. Based on our experience, we outline these architectural requirements, which are essential for building a framework that adapts to the changing nature of CNAs.

\subsubsection{Layered architecture} A layered architecture is a key characteristic desired for effective CNA governance, enabling scalability and controlled management of microservice-based data flows. This approach aligns with the non-functional requirements detailed in Section~\ref{subsec:Desired Non-Functional Requirements}, emphasizing the architecture's adaptability to various complexities of CNA deployments. The significance of employing a layered architecture in CNA governance is further explored in Section~\ref{subsec:Data Flow}, highlighting its crucial role in ensuring flexible and efficient data management throughout the system~\cite{hoque2018architecture, hoque2018online}.

\subsubsection{Adapting well-architected framework principles for governance} Drawing on the foundational principles of the AWS Well-Architected Framework~\cite{AWSWellA87:online} and the IBM Cloud for Financial Services~\cite{IBM-ReferencArchitecture}, it is desired for a reference architecture to integrate these comprehensive guidelines to ensure secure, high-performing, and efficient governance of CNAs. The AWS framework, with its emphasis on operational excellence, security, reliability, performance efficiency, cost optimization, and sustainability~\cite{The6Pill41:online}, offers a versatile set of core concepts applicable across diverse cloud environments. 

Similarly, the IBM Cloud for Financial Services framework focuses on risk management and cloud adoption for regulated industries, incorporating robust security and compliance features~\cite{IBMCloud90:online, Bankingi60:online}. A desired architectural requirement is to adapt and extend these frameworks' principles to a broader governance scope, providing a holistic management perspective for CNAs. Practitioners looking to implement or enhance their governance strategies are advised to explore these frameworks for in-depth guidance.

\subsection{Desired Construction, Deployment, and Maintenance Requirements}\label{subsec:Desired Construction, Deployment, and Maintenance Requirements}

For effective governance across the lifecycle of CNAs, a comprehensive reference architecture must fulfill specific construction, deployment, and maintenance requirements. These requirements, drawn from extensive field experience and inspired by existing cloud-native architectures~\cite{CloudNat67:online, Enabling61:online, ScalingA69:online, goniwada2022cloud}, ensure that the architecture supports a wide range of governance activities effectively. They address the essential aspects of adaptability, resilience, and efficient management crucial for CNAs operating in dynamic cloud environments.

\subsubsection{Vendor agnostic} Essential for broad applicability, the architecture should be designed to be vendor-agnostic, and capable of implementation across all CSPs without dependency on specific proprietary technologies or platforms. 

\subsubsection{Support for multi-cloud} Embracing multi-cloud environments is a desired requirement for enhancing the availability and resilience of CNAs. Recognizing the value of deploying across various CSPs, this architecture aims to support a multi-cloud approach, facilitating a broader range of deployment possibilities and redundancy strategies~\cite{moreno2018orchestrating}. Such flexibility is crucial for tailoring governance mechanisms to leverage the strengths of each CSP effectively. Architectural considerations for data replication and consistency across platforms are essential in multi-cloud setups, ensuring unified governance despite the distributed nature of the resources. This desired feature acknowledges the need for strategic planning in the replication of governance-related data storage, promoting seamless operation across diverse cloud environments.

\subsubsection{Data source and data type agnostic} A crucial desired feature of a governance framework is its ability to operate independently of the specific data sources or types it manages. This agnosticism towards data source and type is fundamental for ensuring the framework's adaptability across various computing environments. Embracing widely recognized data management patterns, such as Publish/Subscribe (Pub/Sub)~\cite{fang2011design}, and Message Queue (MQ)~\cite{tran2011eqs}, enhances this adaptability~\cite{WhatisPu54:online}. These methodologies facilitate efficient data retrieval and dissemination, regardless of the underlying infrastructure, making the architecture versatile and broadly applicable within CNAs.

\subsubsection{XaC and IaC} In an effective governance framework, XaC and IaC are necessary for managing system components through code, improving traceability and automation across cloud computing. IaC, as part of XaC, plays a key role in CNAs management by facilitating seamless delivery and deployment and ensuring high levels of automation~\cite{Architec53:online, WhatisCl52:online, Infrastr92:online}. These approaches offer key benefits such as consistency, scalability, version control, auditability, and portability, critical for various code-as-policy implementations, including IaC, a crucial feature for CNAs~\cite{RealTime91:online, morris2016infrastructure}. This framework leverages XaC and IaC principles, utilizing platforms like GitHub for source code management~\cite{GitHubWh52:online} and tools like Terraform for resource orchestration~\cite{Terrafor55:online} to achieve a comprehensive and automated governance model.
 
\subsubsection{GitOps, DevOps, DevSecOps} GitOps, an operational framework evolved from DevOps practices, has become instrumental in managing CNAs effectively. It extends DevOps and DevSecOps best practices to include XaC, including configurations, infrastructure settings, policies, and deployment constraints. This approach combines IaC with DevOps principles, leading to a fully automated solution delivery model known as GitOps. Central to GitOps is the use of version control systems, like GitHub~\cite{GitHubWh52:online}, as the single source of truth, enabling traceable and reversible states for deployment using CI/CD~\cite{WhatisGi70:online, GuideToG45:online}. CNAs, often built upon microservices and deployed over containers, leverage the latest in CI/CD advancements within the DevOps paradigm~\cite{kratzke2017understanding, Architec53:online, wurster2020cloud}. This setup allows for each microservice to have its own deployment pipeline, enabling simultaneous updates across different CNA components. Adhering to GitOps principles defined by Weaveworks, the entire system of a CNA is described declaratively, with the desired state versioned and changes approved through pull requests or merge requests. Moreover, GitOps' emphasis on observability ensures any divergence from the system's desired state is promptly detected and corrected~\cite{GuideToG45:online, beetz2021gitops}. This comprehensive integration of GitOps and CI/CD practices makes it an apt framework for the development and deployment of CNAs, aligning with current continuous deployment heuristics~\cite{GitOpsGi74:online}.

\subsubsection{End-to-end continuous governance} For effective governance of CNAs across their lifecycle, it is essential to embed a comprehensive, end-to-end governance framework that seamlessly integrates with the Software Development Life Cycle (SDLC). The proposed framework consists of four stages to ensure comprehensive governance.

\begin{itemize}
    \item \textbf{Code-level governance}. This initial stage mandates the continuous review of source code against predefined standards and benchmarks with each update or change request, ensuring code quality and compliance from the outset.
    \item \textbf{CI-level governance}: At this juncture, the focus shifts to verifying the integration of various components within CI pipelines, aligning them with overarching governance goals, and ensuring that the integrated components meet all necessary standards.
    \item \textbf{Pre-CD governance}. Situated at the pivotal intersection of CI and CD, this stage rigorously evaluates component readiness for deployment in diverse environments. It functions as an essential checkpoint, potentially involving a governance team's manual approval to underscore the criticality of meticulous examination prior to automation activation. Successful clearance at this stage triggers the CD pipeline, leading to the deployment of components. 
    \item \textbf{Post-CD governance}. Post-deployment, the governance framework transitions into a maintenance mode, where it actively monitors for any deviations from the predefined `ideal' state, leveraging tools like GitHub as the source of truth. Detected discrepancies trigger corrective actions to realign the current state with the desired one, ensuring continuous compliance and governance integrity.
\end{itemize}

This structured approach to end-to-end continuous governance, underscored by the principle of continuous oversight and adaptation, is designed to navigate CNAs through governance challenges, promoting a unified and strategic governance methodology~\cite{beetz2021gitops}. Many compliance frameworks mandate automated test triggering mechanisms that not only execute predefined tests but also record their results for posterity, ensuring traceability and accountability in compliance verification. This further reinforces the importance of continuous governance in maintaining adherence to regulatory standards.

\subsubsection{Heterogeneous computing environments} Embracing diverse operational demands within CNAs requires the desired adaptability to various computing environments~\cite{scholl2019cloud}. The architecture should ideally support VMs for microservices with continuous operational needs and high resource demands, containers for their scalability and flexibility in microservice architecture, and serverless computing for its efficiency in event-driven and intermittent tasks. Given the continuous advancements in technology, such as Kubernetes' exploration into enhanced GPU management~\cite{Schedule90:online}
, it is crucial for the architecture to adapt and provide governance solutions that are specifically tailored to each computing environment. This approach is aimed at ensuring effective governance across all platforms, thereby enhancing the operational efficiency and security of CNAs within the diverse landscape of cloud computing.

\subsubsection{High availability heartbeat for reliable governance} Ensuring uninterrupted governance in CNAs requires a high availability heartbeat mechanism as a key desired feature. This system continuously monitors the operational health of the governance components, acting as an essential tool for maintaining governance reliability. By identifying potential system failures early and activating necessary failover procedures, the heartbeat mechanism ensures governance operations remain continuous, even in the face of disruptions. This capability is particularly critical in the dynamic and often unpredictable cloud environment, reinforcing the architecture's resilience and adherence to governance standards.

\subsection{Desired Components for Governance of CNAs}\label{subsec:Desired Components for Governance of CNAs}

To achieve comprehensive governance of CNAs, it is necessary to identify desired components within a reference architecture that are crucial for maintaining CNAs' integrity, security, and operational efficiency. Recognizing that some components are inherently composite, combining multiple sub-components to cover various aspects of governance, this discussion provides an in-depth examination of each to clarify their individual and combined contributions. By integrating these desired components, the framework aims to foster a robust and effective governance system, ensuring a resilient and adaptive governance structure tailored to the unique needs of CNAs. \textbf{The list of desired components are}:
\begin{enumerate*}
    \item Data ingestion,
    \item Fault-tolerant data bus,
    \item Data manipulation,
    \item Data storage (mutable and immutable), and
    \item     Data analytics.
\end{enumerate*}
The details are given below.

\subsubsection{Data ingestion} For the comprehensive governance of CNAs, the inclusion of a \component{Data ingestion} component is essential. This component, ideally composite in nature, should consist of \component{Data collector agents} and an \component{API gateway} to address the diverse computing environments of CNAs effectively. \component{Data collector agents} are crucial for fetching and pulling data from varied sources, adapting to heterogeneous computing environments like VMs, containers, and serverless platforms, employing patterns such as the sidecar in VM environments or node-based and pod-based methods in containerized setups. In serverless contexts, where system-level access might be limited, reliance on CSP-provided telemetry or code-level instrumentation becomes necessary for gathering application-specific telemetry. Simultaneously, an \component{API gateway} enhances data ingestion through its dual capability of accepting pushed data and fetching or pulling data, thereby offering a flexible and versatile approach to data collection. This duality is especially critical in CNA environments where data sources are numerous and widely distributed, necessitating adaptable data ingestion strategies. Major CSPs, including AWS~\cite{AWSAPIGateway}, Microsoft Azure~\cite{AzureAPIGateway}, Google Cloud Platform (GCP) ~\cite{GCPAPIGateway}, and IBM Cloud~\cite{IBMAPIGateway}, support this two-way functionality, which is integral for the governance framework, enabling effective monitoring, observability, and compliance. Through the integration of these desired components, the architecture seeks to ensure the thorough and adaptable ingestion of data, a foundation for effective CNA governance.

\subsubsection{Fault tolerant data bus} In the governance of CNAs, a \component{Fault tolerant data bus} is a desired component, essential for managing large volumes of cloud-generated telemetry and facilitating efficient inter-service communication~\cite{pourmajidi2021challenging}. This component, incorporating both a primary and an auxiliary data bus, ensures resilience and continuous operation, which is especially crucial for monitoring, observability, and compliance within CNAs. The primary data bus, typically employing Pub/Sub or MQ models, is supported by major CSPs, including AWS Simple Queue Service~\cite{FullyMan11:online}, Azure Service Bus~\cite{AzureSer41:online}, Google Cloud Pub/Sub~\cite{PubSubfo88:online}, and IBM Cloud Message Queue~\cite{Introduc19:online}. In the event of service disruptions, the auxiliary data bus offers a robust alternative, ensuring that governance processes remain uninterrupted. Designed to operate seamlessly across VMs, containers, and serverless platforms, this dual data bus system employs peer-to-peer networks for enhanced flexibility and resilience, which is critical for the dynamic environments of CNAs. This setup underscores the importance of fault tolerance in maintaining a consistent and reliable governance framework. Gossip-based peer-to-peer networks~\cite{birman2007promise} or more specialized versions for content distribution~\cite{androutsellis2004survey} can be employed, enhancing flexibility and resilience across different cloud scenarios. This strategic approach ensures that governance processes remain continuous and uninterrupted, even in the face of primary data transfer service disruptions. For simplicity, in the rest of this paper, we will refer to these two types of data buses as \component{Data bus}.

\subsubsection{Data manipulation} In the context of effective CNA governance, incorporating \component{Data manipulation} as a desired composite component in a reference architecture is crucial. This component, encompassing \component{Converter, Filter, Aggregator, and Archiver} sub-components, is pivotal for refining and manipulating raw telemetry data for insightful governance analysis. The \component{Converter} ensures data uniformity by transforming diverse telemetry into a standardized format, addressing the challenge posed by the variety of formats. This uniformity is essential for simplifying governance processes and enhancing data analysis efficacy. The \component{Filter's} role is to isolate only the most relevant data for analytics, ensuring efficiency and relevance in data processing. Subsequently, the \component{Aggregator} organizes this filtered data into manageable subsets, facilitating a structured analysis. Finally, the \component{Archiver} is tasked with securely storing these subsets in both immutable and mutable storage options outlined in the \component{Data storage} component. This careful approach to data preservation ensures that governance requirements are consistently met, emphasizing the importance of data integrity and accessibility in the governance framework. 

\subsubsection{Data storage} For the effective governance of CNAs, \component{Data storage} is a desired component. This component, composite in nature, consists of \component{Immutable data storage} and \component{Mutable data storage}. \component{Immutable data storage} plays a critical role in maintaining the integrity and authenticity of digital evidence, which is crucial for compliance and legal frameworks. It is designed to ensure that logs and system-generated telemetry, once written, cannot be altered or deleted. This immutability is highlighted in previous research, pointing to its significance in preserving the admissibility of digital evidence~\cite{pourmajidi2019immutable, pourmajidi2021challenging, reilly2010cloud}. Although blockchain technology presents a robust, tamper-proof option for such storage needs~\cite{pourmajidi2018logchain}, its cost-effectiveness for handling extensive data volumes can be challenging~\cite{chaer2019blockchain, pourmajidi2021immutable}. Alternatively, CSPs have developed Write Once Read Many (WORM) storage solutions, effectively preserving data authenticity without the high costs associated with blockchain~\cite{IBM3363o95:online, WORMAWSS68:online, Overview1:online, Protecti40:online, IBMCloud18:online}.

Conversely, \component{Mutable data storage} addresses a broader spectrum of storage needs within cloud computing, offering flexibility in terms of data access, modification, and deletion. This variety is crucial for governance, allowing for the dynamic management of data that aligns with operational requirements and governance standards. CSPs offer an array of software-defined storage solutions, catering to diverse needs in data type, size, security, and retention policies~\cite{obrutsky2016cloud}. For everyday storage tasks, services like AWS S3~\cite{AmazonS388:online} are suitable~\cite{FreeClou71:online}, while AWS Glacier provides a cost-effective solution for long-term data archiving needs~\cite{AmazonS355:online}. This dual approach to data storage, encompassing both immutable and mutable solutions, is a desired characteristic within the proposed reference architecture, ensuring comprehensive data management that supports the integrity, security, and compliance of CNAs.

While this discussion focuses on fundamental storage options in both mutable and immutable forms, modern storage architectures also include advanced solutions such as data warehouses, data lakes, and data marts. These options offer increased scalability, flexibility, and accessibility, making them suitable for addressing diverse analytical and operational requirements.

\subsubsection{Data analytics} Incorporating pluggable \component{Data analytics} as a desired component within a reference architecture is crucial for ensuring effective governance of CNAs. These components, specifically tailored to address different governance aspects such as monitoring, observability, and compliance, play a significant role in the architecture. Each data analytics component, designed to be atomic and specialized, concentrates on distinct governance requirements, ensuring a seamless fit within the overarching governance framework. For instance, hypothetical examples of data analytics components could include anomaly detection, drift detection, SLA breach detection, Compliance breach detection, and synthetic monitoring, among others. We recommend practitioners develop and plug in their own data analytics components as necessary to cater to their specific governance needs and objectives.

The diversity in the types of assessments conducted by these components—explicit, range, and baseline—demonstrates their comprehensive approach to data analysis. Explicit assessments ensure compliance with predefined criteria, such as verifying the correct number of nodes in a Kubernetes cluster. Range assessments monitor metrics to stay within established limits, like maintaining acceptable database response times. The most complex baseline assessments develop a model of the `desired' state to identify and address deviations, underlining the necessity for adaptive governance~\cite{pourmajidi2019dogfooding, islam2021anomaly, saiful2024anomaly}. These analytics components are indispensable for recognizing deviations from governance standards, including SLAs and configurations, by analyzing data to ascertain if a resource aligns with its defined ideal state. Detected discrepancies prompt immediate actions, enabling proactive governance. Our prior research attests to the efficacy of these data analytics in analyzing cloud-generated data in real-time, underscoring their indispensable role in a governance framework. This adaptability and specialization ensure that the architecture can meet the intricate requirements of cloud governance effectively, highlighting the importance of data analytics as a desired component for comprehensive CNA governance.

\section{Proposed Reference Architecture}\label{sec:Proposed Reference Architecture}

We present our proposed reference architecture in Section~\ref{subsec:Description}. In Section~\ref{subsec:Aligning Governance Components with the Proposed Architecture}, we align the desired governance components (discussed in Section~\ref{subsec:Desired Components for Governance of CNAs}) with the proposed reference architecture. Then, in Section~\ref{subsec:Data Flow}, we provide a detailed overview of the data flow within the reference architecture and among its components. Section~\ref{sec:Architectural Pattern} presents our reference architecture using an architectural pattern template~\cite{bass2012software}. We provide a comparative analysis of our proposed reference architecture in Section~\ref{subsec:Comparative Analysis of Reference Architectures}. Finally, in Section~\ref{sec:Integrating and advancing existing}, we discuss the integration of existing tools with the proposed reference architecture to enhance governance practices.

\subsection{Description}\label{subsec:Description}

CNAs embody the advanced capabilities of cloud technology. Utilizing microservice architectures, APIs, and CI/CD pipelines, CNAs operate within diverse computing environments like containers, serverless frameworks, and VMs. These features render CNAs ideal for enterprise-scale software development. Non-functional requirements such as performance, reliability, and stability are critical, particularly in meeting Quality of Service standards as per SLAs. This is even more pertinent in regulated industries where compliance audits are routine.

The reference architecture we propose integrates governance at its core, offering cloud practitioners, especially in regulated sectors, a blueprint for CNA design and implementation. This integration of governance into the architecture's foundation streamlines the inclusion of governance elements, sparing practitioners the effort of integrating external systems. 
This architecture, developed by experienced cloud professionals and researchers, is designed to handle the complexities of large-scale system monitoring. In particular, a version of this architecture is actively used within IBM Cloud’s observability practices, where it leverages artificial intelligence for near real-time anomaly detection~\cite{pourmajidi2019dogfooding}.

\subsubsection{Core Characteristics}

\begin{figure}[ht]
\centering
\includegraphics[width=\columnwidth]{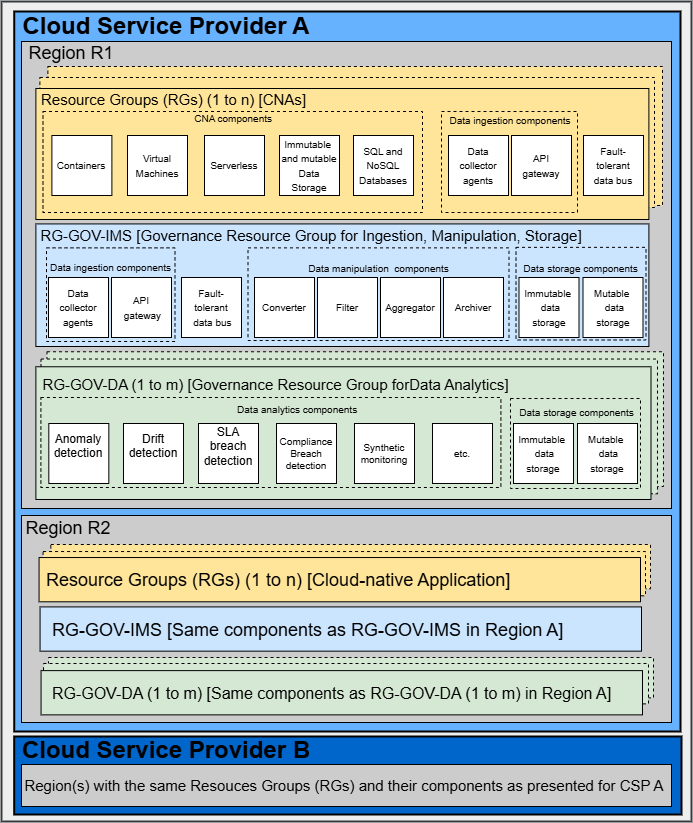}
\caption{Graphical representation of the reference architecture. The CNA and Data Analytics components presented here serve as examples. Practitioners may select components that are most applicable to their specific CNA requirements and data analytics needs.}
\label{fig:reference-architecture}
\end{figure}

The reference architecture is shown in Figure~\ref{fig:reference-architecture}. Our proposed architecture consists of three unique features to address the challenges listed in Section~\ref{subsec:CNA Governance Challenges} as follows. 

\begin{itemize}
    \item \textbf{Multi-vendor cloud support}. The architecture is tailored for multi-vendor cloud environments, enhancing fault tolerance and availability. Figure~\ref{fig:reference-architecture} illustrates this with two CSPs, \csp{CSP A} and \csp{CSP B}, demonstrating the architecture's flexibility and reliability across various cloud services. 
    \item \textbf{Multi-region deployment}. To address geo-redundancy and provide location-aware services, the architecture incorporates multi-region deployment within each CSP. Figure~\ref{fig:reference-architecture} illustrates this with two regions, Region \region{R1} and Region \region{R2}, though the actual number of regions can vary based on service availability and CSP presence in each region.
    \item \textbf{End-to-end governance}. The architecture integrates governance across XaC, IaC, and DevOps processes, leveraging GitOps principles to ensure comprehensive governance coverage. From the initial source code in the Git repository to the final deployment, every stage is designed to meet governance standards, ensuring the process and final solution are fully compliant with governance requirements.
\end{itemize}

\subsubsection{Resource Group Deployment}\label{sec:rg_deployment}

Within each CSP and within each region, multiple resource groups (RGs) are implemented as a means to allocate and manage collections of cloud resources at once. Being a common tool for cloud resource management, RGs are supported by all major CSPs~\cite{AWS-RG,Azure-RG,Google-RG,IBM-RG}. 

\begin{itemize}
    \item \textbf{RGs for CNAs}. 
    Each CNA is placed in an isolated RG with all its operational components. Within RGs $1$ to $n$, we deploy one CNA per RG (i.e., we will have $n$ CNAs in total). For governance practices within CNAs, the architecture separates data ingestion, manipulation, and storage components from data analytics\footnote{Data analytics components are vital for analyzing data in real-time or near-real-time, providing actionable governance insights. Refer to Section~\ref{subsec:Desired Components for Governance of CNAs} for more details.}. This separation allows multiple teams to run their own data analytics on their managed RGs, accessing the same data without duplicating it.
    \item\textbf{Centeralized RG for GOVernance data Ingestion, Manipulation, and Storage (RG-GOV-IMS)}. \RG{RG-GOV-IMS} serves as the centralized hub for consolidating the IMS of telemetry data from all CNAs across RGs $1$ to $n$. This setup ensures that all governance-related data are centrally located, simplifying replication processes, and providing an efficient means for various data analytics components to access the same data. This structure promotes effective governance practices by facilitating the management and analysis of essential governance data.
    \item \textbf{Dedicated RGs for GOVernance Data Analytics (RG-GOV-DA)}. \RG{RG-GOV-DA} serves as a dedicated RG for all types of data analysis against data collected in \RG{RG-GOV-IMS}. Its modular structure allows practitioners to create a dedicated RG for different teams to accommodate their specific governance analytics needs.

\end{itemize}

\subsection{Aligning Governance Components with the Proposed Architecture}\label{subsec:Aligning Governance Components with the Proposed Architecture}

Now that we have explained the flow of data between the internal components of the proposed reference architecture, we can explore the mapping between these components and the desired components for the governance of CNAs, discussed in Section~\ref{subsec:Desired Components for Governance of CNAs}.

The first composite component \component{Data ingestion} is strategically placed in two distinct locations in our reference architecture: within RGs $1$ to $n$ and in \RG{RG-GOV-IMS}. The \component{Data ingestion} components located within RGs $1$ to $n$ are primarily responsible for data collection from the CNAs and their various components. These components are fundamental for gathering telemetry data from various elements of the CNA, ensuring that all relevant information is efficiently captured and relayed. On the other hand, the \component{Data ingestion} component located in \RG{RG-GOV-IMS} is tasked with ingesting data from the \component{Fault tolerant data bus} in RGs $1$ to $n$ into \RG{RG-GOV-IMS}). Their role is central to the data collection that forms the backbone of our governance framework. This setup ensures that data collected from the CNAs is processed and stored, facilitating comprehensive analysis and decision-making. This dual placement of the \component{Data ingestion} component optimizes data flow within our architecture by enabling data collection and ingestion at both the source (CNAs) and the destination (IMS). This strategic positioning of the \component{Data ingestion} components aligns with our overall objective of building an integrated, efficient, and responsive governance system for CNAs.

The second component in our governance architecture is the \component{Fault-tolerant data bus}. This component is deployed within RGs $1$ to $n$ and in \RG{RG-GOV-IMS}. The inclusion of these data buses is crucial for ensuring efficient data flow and governance throughout the entire cloud environment. They serve as critical conduits for transmitting telemetry data and governance-related information. While our architecture diagram distinguishes the data buses in RGs and \RG{RG-GOV-IMS} for clarity, the practical implementation may involve utilizing a singular data bus service, should the CSP offer such a unified solution. For instance, services like AWS Kinesis~\cite{AmazonKi53:online} or IBM Event Streams~\cite{EventStr20:online} could be employed to streamline the data flow processes. This unified approach not only simplifies the data management but also enhances the governance framework's efficiency by reducing the complexity and potential points of failure in the data transmission paths.

The third desired component for robust governance is \component{Data storage}, essential for storing and preserving the authenticity of evidential data. Recognizing its vital role in governance, our proposed reference architecture incorporates \component{Data storage} within both \RG{RG-GOV-IMS}
and \RG{RG-GOV-DA}. This setup ensures that all crucial data remains secure and unaltered, a vital aspect for maintaining governance standards. Moreover, mutable storage offers data storage solutions for data whose authenticity does not need to be proven. These components are implemented in both \RG{RG-GOV-IMS} and \RG{RG-GOV-DA}, utilizing either identical or distinct services provided by CSPs based on specific requirements. This component encompasses a range of storage options, from basic object stores like AWS S3 or IBM COS~\cite{IBMCOS} to more complex NoSQL databases such as AWS DynamoDB~\cite{FastNoSQ6:online} or IBM Cloudant~\cite{IBMCloudant}, catering to the varied data storage demands within a governance framework.

The fourth desired component of our governance-focused architecture is the \component{Data manipulation}, positioned within \RG{RG-GOV-IMS} in our reference architecture. This placement optimally situates this composite component close to both the \component{Fault tolerant data bus} and \component{Data storage} components, allowing more efficient transmission and processing of data crucial for governance. 

The fifth desired component of our reference architecture is the \component{Data analytics}. This composite component consists of many data analytics instances, and they are more than just data analysis tools; rather, they encompass specific business logic aimed at addressing diverse aspects of cloud governance, including monitoring, observability, and compliance. Each \component{Data analytics} component is designed to be focused and atomic, targeting particular governance needs and integrating smoothly into the overall governance framework. To host these \component{Data analytics} components, a dedicated RG, \RG{RG-GOV-DA}
is established. This allocation highlights their importance in identifying deviations from established governance standards and criteria. The data analytics components are essential for analyzing incoming data, ensuring that resources align with their defined desired states, and triggering alerts for proactive governance actions. This design element strengthens the reference architecture's capability to manage the complex and diverse demands of cloud governance.

\subsection{Data Flow}\label{subsec:Data Flow}

Figure~\ref{fig:dataflow} depicts the internal communication among components of the proposed reference architecture. 

\subsubsection{Data bus}
The reference architecture integrates both a primary and auxiliary data bus in \component{Fault tolerant data bus} to strengthen governance robustness. This dual data bus system ensures uninterrupted data flow, a crucial aspect of comprehensive governance. If the primary data bus encounters issues, the auxiliary data bus temporarily takes over, maintaining data transmission until the primary bus is restored (details in Section~\ref{subsec:Desired Components for Governance of CNAs}). This dual-bus approach significantly enhances governance resilience, ensuring continuous data accessibility for governance-related analysis and decision-making.

Additionally, it is important to consider the payload limitations of data buses in a governance context. In scenarios where large objects need to be transmitted but exceed the data bus's payload capacity, a practical solution is to store the payload in a database or cloud storage. The data bus can then exchange pointers to these stored payloads, efficiently handling large object transmission within the governance framework without overloading the data bus capacity~\cite{hoque2018online}. This method ensures that even large-scale data requirements are met within the governance architecture, maintaining the integrity and efficiency of the governance process.

\subsubsection{Data Flow Across RGs}
In line with Section~\ref{sec:rg_deployment}, the architecture utilizes $n$ RGs for hosting CNAs. Within each \RG{RG}, \component{Data ingestion} components are responsible for gathering telemetry data from CNA components. These components employ either \texttt{pull} method~\cite{meehan2017data} by \component{Data collector agents} or \texttt{push} method~\cite{meehan2017data} by \component{API gateway} to collect telemetry from various components of the CNA. Subsequently, \component{Data ingestion} submits its collected telemetry to the \component{Fault tolerant data bus} component within the same \RG{RG}. Given the diverse computing environments employed by CNAs, namely VMs, containers, and serverless platforms, it is essential to deploy \component{Data collector agents} tailored to each environment. This strategic deployment is crucial for effective governance, ensuring that relevant data from various computing platforms are efficiently captured and processed.

\subsubsection{Data Flow in \RG{RG-GOV-IMS}}
In \RG{RG-GOV-IMS}, which is dedicated to data ingestion, manipulation, and storage, the telemetry data from the \component{Fault tolerant data bus} of $n$ RGs are relayed to either the \component{Data collector agents} or the \component{API gateway} components, using either \texttt{pull} or \texttt{push} methods respectively. The choice of method is based on the specific design of each data bus within the RGs.

Once in \RG{RG-GOV-IMS}, telemetry data are passed through \component{Data manipulation} sub-components: \component{Converter}, \component{Filter}, \component{Aggregator}, and \component{Archiver}. These components collectively process the raw telemetry data to make it suitable for governance analysis. The \component{Converter} standardizes the raw data into a unified format, the \component{Filter} isolates data pertinent to data analytics components, the \component{Aggregator} organizes the data into subsets for consumption by data analytics components (see~\cite{hoque2018architecture, hoque2018online} for details), and the \component{Archiver} stores necessary subsets of data into \component{Data storage} composite component. This composite component consists of both \component{Immutable data storage} and \component{Mutable data storage}, with submissions governed by retention policies that reinforce the architecture’s ability to serve CNA governance effectively. The interaction between the \component{Data ingestion} and the \component{Data manipulation} components within \RG{RG-GOV-IMS} is facilitated via its \component{Fault tolerant data bus} component.

\subsubsection{Data Flow in \RG{RG-GOV-DA}}
Within each \RG{RG-GOV-DA}, one or more \component{Data analytics} components are deployed\footnote{Multiple data analytics components often generate alerts that necessitate further analysis and aggregation to discern actual issues and minimize false alerts, thereby reducing "alert fatigue"~\cite{pourmajidi2019dogfooding}. These components can send alerts to a specific topic on the data bus, which are then evaluated by a correlation analytics component. This component identifies genuine alerts, either through predefined rules or machine learning models, creating a bidirectional flow between the \component{Data bus} in \RG{RG-GOV-IMS} and the \component{Pluggable Data Analytics (DA)} in \RG{RG-GOV-DA}.}. Each \component{Data analytics} instance conducts analysis from a distinct governance perspective, converting telemetry data into actionable insights. These insights are then stored within the same \RG{RG-GOV-DA}, either in immutable or mutable storage options available in \component{Data storage} based on compliance requirements for data storage.

Note that \RG{RG-GOV-IMS} serves as the primary data source for all \component{Data analytics} in \RG{RG-GOV-DA} $1$ to $m$. These \component{Data analytics} components access real-time data via the \component{Fault tolerant data bus} and historical data from \component{Data storage} components within \RG{RG-GOV-IMS}, enabling them to perform comprehensive governance analytics.
 
\begin{figure}[t]
\centering
\includegraphics[width=1.0\columnwidth]{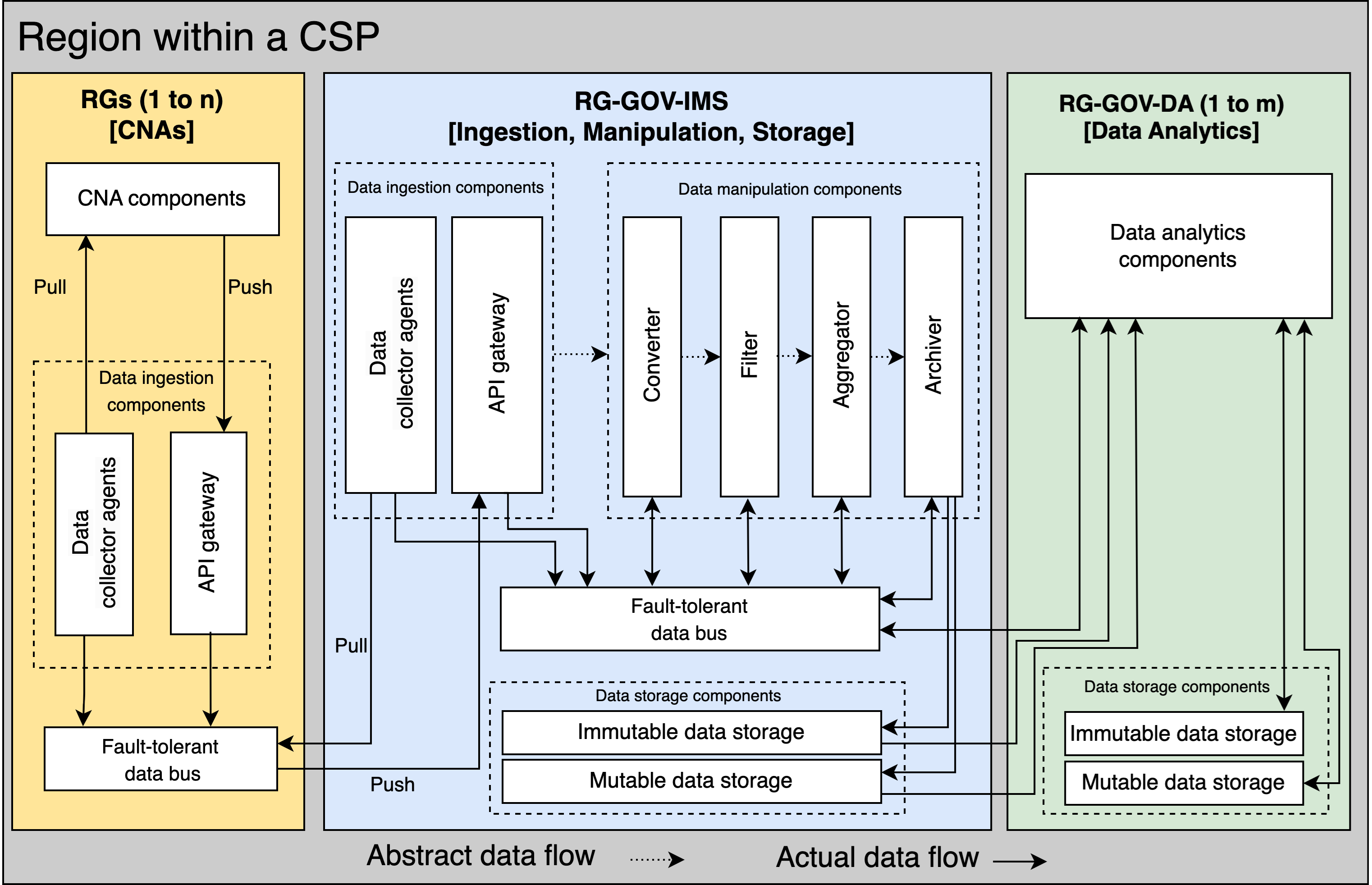}
\caption{Internal communication among components of the proposed reference architecture. Figure~\ref{fig:reference-architecture} shows examples of analytics components in \RG{RG-GOV-DA}. For the sake of brevity, these components are depicted as \component{Data Analytics Components} in this figure. }
\label{fig:dataflow}
\end{figure}

\subsection{Architectural Pattern}\label{sec:Architectural Pattern}

Finally, we summarize information about our reference architecture using an architectural pattern template~\cite[Sec.~13.1]{bass2012software} as follows.

\subsubsection{Context}
CNAs represent an advanced stage in the evolution of application development. These applications, while demonstrating agility and scalability, require robust governance frameworks. Such frameworks extend beyond mere monitoring and observability to include compliance, security, and performance management. Ensuring that CNAs operate reliably and align with both regulatory requirements and organizational goals is essential for their successful deployment and management.

\subsubsection{Problem Description}
CNAs' core characteristics, like heterogeneous computing platforms and microservice architecture, present significant governance challenges. Each computing platform demands specific methods for log collection and analysis. It is crucial to have a detailed view of each component and a holistic view of the entire platform to accurately depict the system's current state. These challenges include managing the distributed nature of microservices, limitations in serverless platforms, and the dynamic aspects of containers. There is also a need to handle diverse telemetry data efficiently, ensuring clean, formatted, and cross-referenced data to reflect an accurate system state. Moreover, governance in CNAs must account for Big Data characteristics inherent in telemetry, demanding advanced solutions for effective storage and analysis. These complexities, detailed in Section~\ref{subsec:CNA Governance Challenges}, underscore the need for a sophisticated and comprehensive governance framework that can address the unique hurdles presented by each architectural element and data type within CNAs.

\subsubsection{Solution Description}\label{subsec:Solution Description}

This reference architecture is tailored for cloud governance, specifically addressing the complexities of CNA governance. It serves as a toolkit for cloud practitioners engaged in designing and implementing CNAs across both regulated and non-regulated industries. Adopting a ``batteries-included'' approach, the architecture provides the necessary components to ensure effective governance within cloud environments. Our extensive and ongoing collaboration with IBM has empirically validated this framework, reinforcing its practical applicability.

\paragraph{Advantages} The architecture, having been applied in real-world CNA scenarios, effectively addresses governance needs. It features a layered, modular design, allowing practitioners to adapt and customize components to fit their unique governance requirements. Being cloud- and technology-neutral, it offers versatility across various cloud services. Based on the latest innovations in cloud computing, as detailed in Section~\ref{subsec:Cloud Computing Evolution}, the architecture supports the creation of new governance frameworks or the enhancement of existing ones (see Section~\ref{sec:Integrating and advancing existing} for details).

\paragraph{Disadvantages} While CNAs commonly utilize cloud technologies, their architectures can vary, often shaped by specific governance needs. Therefore, adopters of this architecture must judiciously assess its suitability for their specific CNA architecture and selectively incorporate the components that align with their governance objectives. While the architecture provides a blueprint of crucial components for governance, it does not offer detailed implementation guidance for each, placing the onus on practitioners to determine the most suitable technologies for each component in their governance framework.

Additionally, while the modular nature of the architecture allows for flexibility, it may introduce integration complexity, particularly when aligning existing tooling with governance pipelines. In high-throughput CNA environments—where telemetry data may be generated at extreme volumes—certain components such as real-time ingestion or analytics layers may become bottlenecks if not scaled appropriately. Moreover, because the architecture offers a generalized reference model rather than prescriptive implementation steps, adopters must invest effort into aligning it with specific compliance regimes, organizational policies, or performance thresholds. This can result in increased time-to-deployment or additional resourcing for architecture customization and validation.

\subsection{Comparative Analysis of Reference Architectures}\label{subsec:Comparative Analysis of Reference Architectures}

Comparative analysis is crucial to evaluate the effectiveness of a reference architecture in meeting established industry standards. We assess the architecture's alignment with key software engineering and architectural criteria that underpin system design and operational excellence in Section~\ref{subsubsec:Software Engineering and Software Architecture Comparison}. In Section~\ref{subsec:Cloud Control Matrix Comparison}, we compare our proposed architecture with the comprehensive governance standards. Through these comparisons, we aim to illustrate our architecture's adherence to industry benchmarks for security, reliability, and compliance, ensuring a governance framework that is both rigorous and resilient.

\subsubsection{Software Engineering and Software Architecture Comparison}\label{subsubsec:Software Engineering and Software Architecture Comparison}

In the realm of CNAs, aligning reference architectures with established software engineering and architectural standards is important. These criteria, or quality attributes, serve as fundamental benchmarks that ensure the architecture not only meets functional requirements but also excels in non-functional dimensions such as maintainability, scalability, and security. Our approach in this comparative analysis is to evaluate how our proposed reference architecture stands against these critical quality attributes, drawing insights from ``Software Architecture in Practice'' by Bass et al.~\cite{bass2012software}. 

To articulate the degree of alignment, we categorize our architecture's support into two distinct classifications: 
\begin{itemize}
    \item \textbf{Fully-supported.} This indicates that the architecture either directly addresses the requirement or provides the flexibility that allows adopters to integrate their solutions seamlessly.

     \item \textbf{Partially-supported.} This denotes cases where the architecture conforms to as many compliance requirements as feasible but cannot fully address certain needs. In such instances, responsibility for implementation rests with the adopter.

\end{itemize}

The proposed architecture demonstrates strong alignment with critical software engineering and architecture quality attributes. A summary of this alignment is provided below, while a detailed breakdown is presented in Appendix~\ref{appsec:SoftwareEngineeringQualityAttributes}, which shows that the architecture fully supports 86\% of the evaluated attributes, addressing essential aspects such as availability, performance, and security. The remaining 14\% are partially supported, requiring some degree of customization or external integration. 

\subsubsection{Cloud Control Matrix Comparison}\label{subsec:Cloud Control Matrix Comparison}

The Cloud Control Matrix (CCM), developed by the Cloud Security Alliance (CSA), is a recognized framework for cloud computing security, offering a detailed guide across 17 domains to ensure comprehensive governance, from risk management to incident response. This framework is essential for organizations utilizing CNAs, serving as a benchmark to assess the security and compliance of cloud services. It has been applied in both academia and industry to evaluate various aspects of cloud governance~\cite{chaudhuri2015governance, saidah2014new, weil2020risk}. 
The proposed architecture aligns with the CCM across multiple governance domains. A summary of this alignment is provided below, with a detailed evaluation presented in Appendix~\ref{appsec:CCMAlignment}, which reveals that 65\% of the evaluated domains are fully supported, covering critical areas such as audit, encryption, and governance. Additionally, 24\% are partially supported, requiring some degree of customization, while 12\% are out of scope.

\subsection{Integrating and Advancing Existing Solutions for Enhanced Cloud Governance}\label{sec:Integrating and advancing existing}
Incorporating established tools such as Elasticsearch~\cite{ElasticsSearch}, Logstash~\cite{Logstash}, Kibana~\cite{Kibana}, Mezmo~\cite{Mezmo}, Splunk~\cite{Splunk}, and Sysdig~\cite{Sysdig} within our governance framework is a strategic approach, leveraging their unique capabilities to enhance functionality. Our architecture emphasizes leveraging these existing resources, steering clear of the complexities involved in developing components from scratch. By channeling telemetry data to \component{Data collector agents} in \RG{
RG-GOV-IMS}, we facilitate seamless integration of telemetry and alerts into a unified RG. To ensure consistency across diverse telemetry formats, we employ \component{Data manipulation} sub-components for standardization.

Many governance tools provide standardized mechanisms to either push data (e.g., through webhooks or streaming APIs) or allow periodic data pulls (e.g., via scheduled queries or REST APIs). These mechanisms are typically backed by persistent storage layers, including relational databases, NoSQL engines, or other storage systems. The proposed architecture can accommodate these data exchange strategies by aligning them with the ingestion pathways defined in the \component{Data collector agents} and \component{Data manipulation} stages within the \RG{RG-GOV-IMS} resource group. These integration points make it feasible to import telemetry, audit trails, or configuration data from a variety of third-party tools, providing both observability and compliance enrichment while avoiding redundant tooling or duplicated pipelines.

Ultimately, adopters of our reference architecture can efficiently instantiate additional instances using IaC, reducing the time and effort required compared to building an original implementation. This approach enhances the governance architecture's adaptability and scalability and ensures a more resource-efficient deployment.

\section{Real-world Instantiation of the Proposed Reference Architecture}\label{sec:Real-world Instantiation of the Proposed Reference Architecture}

We implemented the proposed reference architecture, emphasizing its adaptability and scalability across multi-cloud environments. The deployment demonstrates the functionality of resource groups in handling telemetry and log data from high-demand and less-demanding environments, showcasing key architectural elements. The implementation includes:

A representative subset of the architecture is deployed, with AWS hosting a high-demand CNA environment and IBM Cloud hosting a less-demanding CNA environment. Governance components (\RG{RG-GOV-IMS} and \RG{RG-GOV-DA}) are implemented on AWS for consistency.

CNAs are containerized using Docker and deployed on both AWS and IBM. These CNAs emit telemetry data, including metrics and logs, every second for a total of 4,000 entries. AWS employs additional components, such as API Gateways and SQS, to handle the high velocity of data, while IBM utilizes a streamlined approach to transmit telemetry directly to governance pipelines.

The \RG{RG-GOV-IMS} resource group processes telemetry data through a structured pipeline, ensuring fault tolerance, scalability, and compliance. It employs API Gateways, SNS topics, and Lambda functions for ingestion, manipulation, and storage. Immutable and mutable S3 storage options are used for logs and metrics, respectively, with object-lock enabled for regulatory adherence.

The \RG{RG-GOV-DA} resource group analyzes telemetry data for performance evaluation. It measures time delays across pipeline stages, comparing AWS and IBM implementations. Insights include variations in delay times due to inter-regional data transfer and internal architecture differences.
											
We demonstrate the system’s ability to maintain performance consistency across cloud providers.  A graphical representation of the data flow within the instantiated reference architecture is available in Appendix~\ref{appsec:Real-world Instantiation of the Proposed Reference Architecture}, providing a visual walkthrough of the end-to-end operational workflow.																				
This deployment also serves as a practical case study to guide cloud and cloud-native adopters in implementing governance architectures tailored to their own environments. Recognizing that no two CNA deployments are alike, the proposed framework is intentionally designed for flexibility, allowing organizations to incorporate additional components or modify existing ones to meet their specific needs. 
		   
For comprehensive implementation details, including deployment instructions, telemetry formats, data flow, time-delay analytics, and cross-cloud performance metrics, refer to Appendix~\ref{appsec:Real-world Instantiation of the Proposed Reference Architecture}. 

The GitHub repository~\cite{WilliamPCNA_Governance_Git:online} contains example code, deployment scripts, and documentation structured around GitOps principles, facilitating modular, reproducible multi-cloud deployments.

\section{Related Literature}\label{sec:Related Literature}

This section explores the literature related to CNAs, covering monitoring, observability, compliance, and governance. We discuss CNA monitoring, observability, and compliance in Section~\ref{subsec:CNA Monitoring, Observability, and Compliance}, highlighting the need for integrated approaches in these areas. Section~\ref{subsec:CNA Governance litreview}
 focuses on CNA governance, while Section~\ref{subsec:Existing Tools and Practices in Governance of CNAs} examines existing tools and practices. 

Given the interdisciplinary nature of this work and the wide range of adjacent research domains, we have limited our literature review to peer-reviewed studies that focus specifically on CNA-related monitoring, observability, compliance, and governance. This targeted scope ensures relevance to the core components of our proposed architecture and avoids diluting the discussion with broader cloud or software engineering topics not directly aligned with CNA governance.

\subsection{CNA Monitoring, Observability, and Compliance}\label{subsec:CNA Monitoring, Observability, and Compliance}
Monitoring CNAs is a complex task due to their deployment over elastic platforms, utilization of various computing environments (VMs, containers, serverless), and microservice architectures. Both detailed and unified views of applications are essential for the effective governance and maintenance of CNAs.

Brunner et al.~\cite{brunner2015experimental} highlight that CNAs inherently support self-management, scalability, and resilience across clustered units of application logic. However, there is no dominant application development model for CNAs, presenting challenges in establishing comprehensive governance models. This gap is where our work aims to make a significant contribution.

Oliveira et al.~\cite{oliveira2017cloud} stress the importance of operational visibility in cloud services' success. They point out the complexities in monitoring cloud environments and the need for adaptable solutions to the diverse and changing nature of cloud resources. Their work on a unified monitoring and analytics pipeline, OpVis, provides insights into addressing some of these challenges, although focusing primarily on the observability aspect of CNAs.

However, comprehensive governance (including compliance, security, and policy adherence)  is not addressed in~\cite{brunner2015experimental,oliveira2017cloud}. Our proposed reference architecture (see Section~\ref{sec:Proposed Reference Architecture}) aims to bridge this gap by providing a holistic governance framework that encompasses monitoring, observability, compliance, and security, thus ensuring robust and efficient management of CNAs.

Torkura et al.~\cite{torkura2017integrating} shed light on the compliance challenges specific to CNAs. They underscore CNAs' vulnerability due to the diversity of technologies used and rapid development cycles. They highlight the inadequacy of traditional security assessments in dynamic CNAs, where microservices are frequently created and terminated. To address these issues, they propose new approaches like the Security Gateway, dynamic document store, and security health endpoints integrated into the CNA workflow. Torkura et al. indicate that existing research primarily focuses on security mechanisms such as authentication, authorization, and network security, leaving a gap in comprehensive vulnerability detection and security assessments.

Our work aligns with the insights of Torkura et al., recognizing the need for a governance framework that goes beyond monitoring and observability to include compliance. We aim to fill the gaps identified by Torkura et al., embedding compliance within our reference architecture. This approach ensures that CNAs not only meet operational requirements but also comply with security standards and regulatory demands. Our architecture, therefore, provides a comprehensive solution to the complex compliance landscape of CNAs, where existing research is insufficient.

\subsection{CNA Governance}\label{subsec:CNA Governance litreview}
In their study, Laszewski et al.~\cite{laszewski2018cloud} explore the evolution of cloud computing governance. They emphasize that cloud computing offers superior security and governance compared to traditional data center operations. Major cloud vendors continuously integrate new security patterns and compliance requirements into their services, ensuring customers benefit from enhanced security and governance. By using cloud services, organizations automatically align with these evolving standards, often without incurring additional costs. However, leveraging cloud security effectively requires meticulous implementation and ongoing governance. CSPs offer a wide range of services to address various security needs, but it is the responsibility of individual CSP to implement these services properly. Maintaining and monitoring the implemented security measures demands substantial governance efforts. Cloud services offer tools for monitoring, auditing, and configuration control, creating an overlay of governance activities. This unified cloud governance model ensures consistency across deployed workloads, avoiding gaps that may arise in on-premises deployments due to complexity or cost.  The authors also highlight various governance guidelines designed to support secure and compliant operations in the cloud. These guidelines cater to different industry verticals, data requirements, and external compliance needs, offering pathways to enhance cloud security postures. Even for organizations without specific industry compliance requirements, adopting these governance guidelines can significantly enhance the resilience and security of cloud resources, supporting internal audit and control processes. Laszewski et al. emphasize the importance of these governance models in ensuring strict adherence to security and compliance standards in cloud environments, a critical factor for effective CNA governance.

Kosinska et al.~\cite{kosinska2020autonomic} delve into governance for cloud-hosted applications, including CNAs. They compare cloud computing to an advanced form of service-oriented computing (SOC), where cloud platforms offer not just data and applications but also a range of computing resources as services. This characteristic makes cloud-hosted applications act like services, often depending on other cloud applications or interacting with remote clients via APIs. This similarity with SOC suggests that many SOA governance concepts are directly applicable to cloud platforms and cloud-hosted applications. The authors define governance for cloud-hosted applications as ensuring these applications operate as expected while maintaining a certain quality of service. They focus on three governance features: policy enforcement, formulating performance SLOs, and application performance monitoring (APM). Policy enforcement involves ensuring compliance with various standards and practices, such as dependency management and software versioning requirements. Performance SLOs are statistical bounds on application performance used to negotiate SLAs and monitor application consistency. APM involves continuous monitoring of cloud-hosted applications to detect and diagnose performance anomalies. However, current cloud technologies do not satisfactorily implement these governance features, leading to the emergence of third-party governance solutions. These external services, while offering access control and API management, are often expensive and require additional programming or configuration. They also suffer from limited visibility and control over applications residing in the cloud. Kosinska et al. argue for built-in CNA governance capabilities, which they demonstrate as more robust, effective, and user-friendly compared to external solutions. Furthermore, they highlight the rapid proliferation of web-accessible APIs and the need for automated API governance. Modern applications increasingly rely on remote web APIs, reducing programming and maintenance workloads. This growing reliance on APIs calls for new techniques in API governance, focusing on consistent policy implementation across multiple APIs. Kosinska et al.'s research targets API governance in cloud-hosted web applications, proposing systems for configuring and enforcing policies at the API level and stipulating performance SLAs for individual APIs. This study by Kosinska et al. contributes to our understanding of cloud governance, particularly in the context of cloud-hosted applications and APIs. It underscores the need for integrated CNA governance mechanisms, highlighting the limitations of current approaches and advocating for more robust solutions that are in tune with the unique demands of cloud environments.

\subsection{Existing Tools and Practices in Governance of CNAs}\label{subsec:Existing Tools and Practices in Governance of CNAs}
Toka et al.~\cite{toka2021predicting} highlight the significance of fault detection in Cloud applications. They propose an open-source, CNA monitoring solution that emphasizes the need for lightweight yet effective tools in dynamic cloud environments. Their solution integrates elements such as collectors, databases, and visualization engines, underscoring the need for comprehensive monitoring frameworks.

Wang et al.~\cite{wang2018cloudranger} introduce CloudRanger, a novel root cause analysis tool designed for CNAs. CloudRanger's ability to adapt to the dynamic nature of cloud infrastructures exemplifies the complexity of monitoring cloud applications. The tool's unique approach to establishing dynamic, causal relationships and constructing impact graphs for applications without a fixed topology is particularly noteworthy.

These monitoring tools and methodologies, while effective, typically focus on isolated aspects of CNAs, such as fault detection or root cause analysis. However, in the broader context of cloud governance, there is a need for an integrated approach that encompasses not only monitoring and observability but also compliance, security, and policy management. Our proposed reference architecture (see Section~\ref{sec:Proposed Reference Architecture}) aims to fill this gap. By integrating existing tools and practices with additional governance components, we provide a holistic framework for managing CNAs. This integrated approach ensures that CNAs are not only monitored for performance and reliability but also governed in line with compliance requirements and security best practices, ultimately enhancing the overall management and operational efficiency of CNAs.

\section{Conclusion}\label{sec:Conclusion}
This paper explores the realm of CNAs and the challenges involved in governing them. We present a reference architecture for robust CNA governance that is applicable across a range of single and multi-cloud environments. Rather than proposing a one-size-fits-all model, the architecture is designed to be modular and adaptable, enabling integration with diverse deployment contexts and enterprise governance policies. By embedding governance directly into CNA design, the framework offers a ``batteries-included'' approach that streamlines compliance efforts and enables cloud professionals to focus on core product development. The architecture’s real-world instantiation illustrates its flexibility and demonstrates how it may be tailored to meet specific organizational, regulatory, or operational needs. While this work highlights broad applicability, we also recognize that cloud environments vary in scale, tooling, and governance maturity. As such, additional customization may be required in some contexts. Overall, the proposed architecture aims to improve CNA management and operational efficiency, serving as a practical foundation for further research and standardization in cloud-native governance.

Integrating AI and ML into CNA governance frameworks can automate policy enforcement and compliance, transforming regulatory requirements into executable code to enhance consistency and reduce human error. AI-driven predictive analytics can proactively identify compliance risks, enabling timely remediation. Future research should focus on seamlessly incorporating these technologies into governance pipelines to maintain scalability and robustness across diverse cloud environments.

\section*{Acknowledgments}
This work was partially supported by the Natural Sciences and Engineering Research Council of Canada (Grant \# RGPIN-2022-03886) and the IBM Centre for Advanced Studies (Project \#1142).

\bibliographystyle{IEEEtran}
\bibliography{reference.bib}

\begin{thebibliography}{100}
\providecommand{\url}[1]{#1}
\csname url@samestyle\endcsname
\providecommand{\newblock}{\relax}
\providecommand{\bibinfo}[2]{#2}
\providecommand{\BIBentrySTDinterwordspacing}{\spaceskip=0pt\relax}
\providecommand{\BIBentryALTinterwordstretchfactor}{4}
\providecommand{\BIBentryALTinterwordspacing}{\spaceskip=\fontdimen2\font plus
\BIBentryALTinterwordstretchfactor\fontdimen3\font minus \fontdimen4\font\relax}
\providecommand{\BIBforeignlanguage}[2]{{%
\expandafter\ifx\csname l@#1\endcsname\relax
\typeout{** WARNING: IEEEtran.bst: No hyphenation pattern has been}%
\typeout{** loaded for the language `#1'. Using the pattern for}%
\typeout{** the default language instead.}%
\else
\language=\csname l@#1\endcsname
\fi
#2}}
\providecommand{\BIBdecl}{\relax}
\BIBdecl

\bibitem{IDCFutureScape:online}
``Idc-futurescape,'' \url{http://phc.pt/enews/IDC-FutureScape.pdf}, 2020.

\bibitem{hsu2014examining}
P.-F. Hsu, S.~Ray, and Y.-Y. Li-Hsieh, ``Examining cloud computing adoption intention, pricing mechanism, and deployment model,'' \emph{International Journal of Information Management}, vol.~34, no.~4, pp. 474--488, 2014.

\bibitem{nandgaonkar2014comprehensive}
S.~V. Nandgaonkar and A.~Raut, ``A comprehensive study on cloud computing,'' \emph{International Journal of Computer Science and Mobile Computing}, vol.~3, no.~4, pp. 733--738, 2014.

\bibitem{NISTClou61:online}
``Nist cloud computing standards roadmap,'' \url{https://nvlpubs.nist.gov/nistpubs/SpecialPublications/NIST.SP.500-291r2.pdf}, (Accessed on 12/08/2024).

\bibitem{hogan2011nist}
M.~Hogan, F.~Liu \emph{et~al.}, ``Nist cloud computing standards roadmap,'' \emph{NIST Special Publication}, vol.~35, pp. 6--11, 2011.

\bibitem{redhat-cloudgovdefinition}
``What is cloud governance?'' \url{https://www.redhat.com/en/topics/automation/what-is-cloud-governance}, Jan 2023.

\bibitem{AWSCloud85:online}
AWS, ``Aws\_cloudgovernance\_ebook\_driving-success-and-security-in-the-cloud.pdf,'' \url{https://pages.awscloud.com/rs/112-TZM-766/images/AWS_CloudGovernance_ebook_Driving-Success-and-Security-in-the-Cloud.pdf}, 2021.

\bibitem{Governme20:online}
``Govern methodology for the cloud - cloud adoption framework | microsoft learn,'' \url{https://learn.microsoft.com/en-ca/azure/cloud-adoption-framework/govern/methodology}, 5 2025.

\bibitem{WhatisDa76:online}
``What is data governance?  |  google cloud,'' \url{https://cloud.google.com/learn/what-is-data-governance}, 2025.

\bibitem{IBMCloud51:online}
``Ibm cloud compliance overview page,'' \url{https://www.ibm.com/cloud/compliance}, 2025.

\bibitem{Whatisda59:online}
``What is data governance? | ibm,'' \url{https://www.ibm.com/topics/data-governance}, 9 2024.

\bibitem{varghese2018next}
B.~Varghese and R.~Buyya, ``Next generation cloud computing: New trends and research directions,'' \emph{Future Generation Computer Systems}, vol.~79, pp. 849--861, 2018.

\bibitem{abbasi2019software}
A.~A. Abbasi, A.~Abbasi \emph{et~al.}, ``Software-defined cloud computing: A systematic review on latest trends and developments,'' \emph{IEEE Access}, vol.~7, pp. 93\,294--93\,314, 2019.

\bibitem{WhatisaC94:online}
``What is a container?'' \url{https://www.docker.com/resources/what-container}, Docker, 2025.

\bibitem{liu2020microservices}
G.~Liu, B.~Huang \emph{et~al.}, ``Microservices: architecture, container, and challenges,'' in \emph{2020 IEEE 20th international conference on software quality, reliability and security companion (QRS-C)}.\hskip 1em plus 0.5em minus 0.4em\relax IEEE, 2020, pp. 629--635.

\bibitem{Kubernet86:online}
``Kubernetes,'' \url{https://kubernetes.io/}, 2025.

\bibitem{van2018serverless}
E.~Van~Eyk, L.~Toader \emph{et~al.}, ``Serverless is more: From paas to present cloud computing,'' \emph{IEEE Internet Computing}, vol.~22, no.~5, pp. 8--17, 2018.

\bibitem{CloudNat91:online}
``Cloud native apps; cloud native glossary,'' \url{https://glossary.cncf.io/cloud-native-apps/}, 2025.

\bibitem{odun2019systematic}
I.~Odun-Ayo, R.~Goddy-Worlu \emph{et~al.}, ``A systematic mapping study of cloud-native application design and engineering,'' in \emph{Journal of Physics: Conference Series}, vol. 1378, no.~3.\hskip 1em plus 0.5em minus 0.4em\relax IOP Publishing, 2019, p. 032092.

\bibitem{alonso2019decide}
J.~Alonso, K.~Stefanidis \emph{et~al.}, ``Decide: an extended devops framework for multi-cloud applications,'' in \emph{3rd international conference on cloud and big data computing}, 2019, pp. 43--48.

\bibitem{alonso2023understanding}
J.~Alonso, L.~Orue-Echevarria \emph{et~al.}, ``Understanding the challenges and novel architectural models of multi-cloud native applications--a systematic literature review,'' \emph{Journal of Cloud Computing}, vol.~12, no.~1, p.~6, 2023.

\bibitem{9284234}
B.~Thuraisingham, ``Cloud governance,'' in \emph{2020 IEEE 13th Int. Conf. on Cloud Computing (CLOUD)}, 2020, pp. 86--90.

\bibitem{Terrafor55:online}
``Terraform,'' \url{https://www.terraform.io/}, HashiCorp.

\bibitem{ChefSoft16:online}
``Chef software devops automation solutions,'' \url{https://www.chef.io/}, Chef, 2025.

\bibitem{Infrastr7:online}
``Infrastructure automation with ansible,'' \url{https://www.ansible.com/integrations/infrastructure?hsLang=en-us}, Ansible, 2025.

\bibitem{leymann2016native}
F.~Leymann, U.~Breitenb{\"u}cher \emph{et~al.}, ``Native cloud applications: why monolithic virtualization is not their foundation,'' in \emph{Int. Conf. on Cloud Computing and Services Science}.\hskip 1em plus 0.5em minus 0.4em\relax Springer, 2016, pp. 16--40.

\bibitem{kratzke2017understanding}
N.~Kratzke and P.-C. Quint, ``Understanding cloud-native applications after 10 years of cloud computing-a systematic mapping study,'' \emph{Journal of Systems and Software}, vol. 126, pp. 1--16, 2017.

\bibitem{balalaie2015migrating}
A.~Balalaie, A.~Heydarnoori, and P.~Jamshidi, ``Migrating to cloud-native architectures using microservices: an experience report,'' in \emph{European Conference on Service-Oriented and Cloud Computing}.\hskip 1em plus 0.5em minus 0.4em\relax Springer, 2015, pp. 201--215.

\bibitem{Microser77:online}
M.~Fowler, ``Microservices,'' \url{https://martinfowler.com/articles/microservices.html}, 3 2014.

\bibitem{zhang2019microservice}
H.~Zhang, S.~Li \emph{et~al.}, ``Microservice architecture in reality: An industrial inquiry,'' in \emph{2019 IEEE international conference on software architecture (ICSA)}.\hskip 1em plus 0.5em minus 0.4em\relax IEEE, 2019, pp. 51--60.

\bibitem{singh2017container}
V.~Singh and S.~K. Peddoju, ``Container-based microservice architecture for cloud applications,'' in \emph{2017 Int. Conf. on Computing, Communication and Automation (ICCCA)}.\hskip 1em plus 0.5em minus 0.4em\relax IEEE, 2017, pp. 847--852.

\bibitem{amaral2015performance}
M.~Amaral, J.~Polo \emph{et~al.}, ``Performance evaluation of microservices architectures using containers,'' in \emph{2015 IEEE 14th Int. Symp. on Network Computing and Applications}.\hskip 1em plus 0.5em minus 0.4em\relax IEEE, 2015, pp. 27--34.

\bibitem{ManagedK42:online}
``Managed kubernetes service,'' \url{https://aws.amazon.com/eks/}, AWS, 2025.

\bibitem{ManagedK46:online}
``Managed kubernetes service (aks),'' \url{https://azure.microsoft.com/en-us/services/kubernetes-service/#overview}, Microsoft, 2024.

\bibitem{Kubernet19:online}
``Kubernetes - google kubernetes engine (gke),'' \url{https://cloud.google.com/kubernetes-engine}, Google, 2025.

\bibitem{li2009method}
X.~Li, Y.~Li \emph{et~al.}, ``The method and tool of cost analysis for cloud computing,'' in \emph{2009 IEEE Int. Conf. on Cloud Computing}.\hskip 1em plus 0.5em minus 0.4em\relax IEEE, 2009, pp. 93--100.

\bibitem{eivy2017wary}
A.~Eivy and J.~Weinman, ``Be wary of the economics of" serverless" cloud computing,'' \emph{IEEE Cloud Computing}, vol.~4, no.~2, pp. 6--12, 2017.

\bibitem{castro2019rise}
P.~Castro, V.~Ishakian \emph{et~al.}, ``The rise of serverless computing,'' \emph{Communications of the ACM}, vol.~62, no.~12, pp. 44--54, 2019.

\bibitem{taibi2020serverless}
D.~Taibi, J.~Spillner, and K.~Wawruch, ``Serverless computing-where are we now, and where are we heading?'' \emph{IEEE Software}, vol.~38, no.~1, pp. 25--31, 2020.

\bibitem{WhatareC56:online}
``What are cloud native applications?'' \url{https://tanzu.vmware.com/cloud-native}, VMware, 2025.

\bibitem{Serverle63:online}
``Serverless computing - aws lambda,'' \url{https://aws.amazon.com/lambda/}, AWS, 2025.

\bibitem{AzureSer62:online}
``Azure serverless,'' \url{https://azure.microsoft.com/en-ca/solutions/serverless/}, Microsoft, 2025.

\bibitem{Serverle17:online}
``Serverless computing,'' \url{https://cloud.google.com/serverless}, Google, 2025.

\bibitem{IBMCloud93:online}
``Ibm cloud code engine - canada,'' \url{https://www.ibm.com/ca-en/cloud/code-engine}, IBM, 2025.

\bibitem{wurster2020cloud}
M.~Wurster, U.~Breitenb{\"u}cher \emph{et~al.}, ``Cloud-native deploy-ability: An analysis of required features of deployment technologies to deploy arbitrary cloud-native applications.'' in \emph{CLOSER}, 2020, pp. 171--180.

\bibitem{fehling2014cloud}
C.~Fehling, F.~Leymann \emph{et~al.}, \emph{Cloud computing patterns: fundamentals to design, build, and manage cloud applications}.\hskip 1em plus 0.5em minus 0.4em\relax Springer, 2014.

\bibitem{levin2020viperprobe}
J.~Levin and T.~A. Benson, ``Viperprobe: Rethinking microservice observability with ebpf,'' in \emph{2020 IEEE 9th Int. Conf. on Cloud Networking (CloudNet)}.\hskip 1em plus 0.5em minus 0.4em\relax IEEE, 2020, pp. 1--8.

\bibitem{moradi2017conmon}
F.~Moradi, C.~Flinta \emph{et~al.}, ``Conmon: An automated container based network performance monitoring system,'' in \emph{2017 IFIP/IEEE Symp. on Integrated Network and Service Management (IM)}.\hskip 1em plus 0.5em minus 0.4em\relax IEEE, 2017, pp. 54--62.

\bibitem{raj2018automated}
P.~Raj, A.~Raman \emph{et~al.}, ``Automated multi-cloud operations and container orchestration,'' \emph{Software-Defined Cloud Centers: Operational and Management Technologies and Tools}, pp. 185--218, 2018.

\bibitem{martin2021observability}
P.~Martin, ``Observability,'' in \emph{Kubernetes}.\hskip 1em plus 0.5em minus 0.4em\relax Springer, 2021, pp. 175--183.

\bibitem{cordingly2021enhancing}
R.~Cordingly, N.~Heydari \emph{et~al.}, ``Enhancing observability of serverless computing with the serverless application analytics framework,'' in \emph{Companion of the ACM/SPEC Int. Conf. on Performance Engineering}, 2021, pp. 161--164.

\bibitem{gatev2021observability}
R.~Gatev, ``Observability: Logs, metrics, and traces,'' in \emph{Introducing Distributed Application Runtime (Dapr)}.\hskip 1em plus 0.5em minus 0.4em\relax Springer, 2021, pp. 233--252.

\bibitem{miranskyy2007iterative}
A.~V. Miranskyy, N.~H. Madhavji \emph{et~al.}, ``An iterative, multi-level, and scalable approach to comparing execution traces,'' in \emph{6th joint meeting of the European software engineering conference and the ACM SIGSOFT symposium on the foundations of software engineering}, 2007, pp. 537--540.

\bibitem{miranskyy2008sift}
------, ``Sift: a scalable iterative-unfolding technique for filtering execution traces,'' in \emph{Conference of the center for advanced studies on collaborative research: meeting of minds}, 2008, pp. 274--288.

\bibitem{MetricsL15:online}
``Metrics, logs and traces: The golden triangle of observability in monitoring,'' \url{https://devops.com/metrics-logs-and-traces-the-golden-triangle-of-observability-in-monitoring/}, 11 2018.

\bibitem{miranskyy2016operational}
A.~Miranskyy, A.~Hamou-Lhadj \emph{et~al.}, ``Operational-log analysis for big data systems: Challenges and solutions,'' \emph{IEEE Software}, vol.~33, no.~2, pp. 52--59, 2016.

\bibitem{pourmajidi2017challenges}
W.~Pourmajidi, J.~Steinbacher \emph{et~al.}, ``On challenges of cloud monitoring,'' in \emph{27th Annual Int. Conf. on Computer Science and Software Engineering}, 2017, pp. 259--265.

\bibitem{islam2020anomaly}
M.~S. Islam and A.~Miranskyy, ``Anomaly detection in cloud components,'' in \emph{IEEE 13th Int. Conf. on Cloud Computing (CLOUD)}, 2020, pp. 1--3.

\bibitem{islam2021anomaly}
M.~S. Islam, W.~Pourmajidi \emph{et~al.}, ``Anomaly detection in a large-scale cloud platform,'' in \emph{2021 IEEE/ACM 43rd Int. Conf. on Software Engineering: Software Engineering in Practice (ICSE-SEIP)}.\hskip 1em plus 0.5em minus 0.4em\relax IEEE, 2021, pp. 150--159.

\bibitem{pourmajidi2019dogfooding}
W.~Pourmajidi, A.~Miranskyy \emph{et~al.}, ``Dogfooding: Using ibm cloud services to monitor ibm cloud infrastructure,'' in \emph{29th Annual Int. Conf. on Computer Science and Software Engineering}, 2019, pp. 344--353.

\bibitem{pourmajidi2021challenging}
W.~Pourmajidi, L.~Zhang \emph{et~al.}, ``The challenging landscape of cloud monitoring,'' in \emph{Knowledge Management in the Development of Data-Intensive Systems}.\hskip 1em plus 0.5em minus 0.4em\relax CRC Press, 2021, pp. 157--189.

\bibitem{sohana2024cloudheatmap}
S.~Sohana, W.~Pourmajidi \emph{et~al.}, ``Cloudheatmap: Heatmap-based monitoring for large-scale cloud systems,'' \emph{arXiv preprint arXiv:2410.21092}, 2024.

\bibitem{saiful2024anomaly}
M.~Saiful~Islam, M.~Sami~Rakha \emph{et~al.}, ``Anomaly detection in large-scale cloud systems: An industry case and dataset,'' \emph{arXiv e-prints}, pp. arXiv--2411, 2024.

\bibitem{WhatIsCl93:online}
``What is cloud native observability?'' \url{https://newrelic.com/blog/best-practices/what-is-cloud-native-observability#:~:text=Observability%20is%20defined%20as%20your,data%20processes%2C%20and%20hardware%20processes.}, New Relic, 2 2024.

\bibitem{esposito2016challenges}
C.~Esposito, A.~Castiglione, and K.-K.~R. Choo, ``Challenges in delivering software in the cloud as microservices,'' \emph{IEEE Cloud Computing}, vol.~3, no.~5, pp. 10--14, 2016.

\bibitem{yu2016cloudseer}
X.~Yu, P.~Joshi \emph{et~al.}, ``Cloudseer: Workflow monitoring of cloud infrastructures via interleaved logs,'' \emph{ACM SIGARCH Computer Architecture News}, vol.~44, no.~2, pp. 489--502, 2016.

\bibitem{Summaryo12:online}
``Summary of the amazon s3 service disruption in the northern virginia (us-east-1) region,'' \url{https://aws.amazon.com/message/41926/}, AWS, 2017.

\bibitem{AmazonAn9:online}
``Amazon and the \$150 million typo : The two-way,'' \url{https://www.npr.org/sections/thetwo-way/2017/03/03/518322734/amazon-and-the-150-million-typo}, NPR, 3 2017.

\bibitem{wahab-aiops}
A.~Hamou-Lhadj, ``{A Call for the Development of an International Standard for the Management of AIOps Systems},'' in \emph{ASPLOS Workshop on Cloud Intelligence and AIOps}, 2024, p.~1.

\bibitem{bohner2012controllability}
M.~Bohner and N.~Wintz, ``Controllability and observability of time-invariant linear dynamic systems,'' \emph{Mathematica Bohemica}, vol. 137, no.~2, pp. 149--163, 2012.

\bibitem{kalman1960general}
R.~E. Kalman, ``On the general theory of control systems,'' in \emph{Proceedings First Int. Conf. on Automatic Control, Moscow, USSR}, 1960, pp. 481--492.

\bibitem{Observability}
R.~Picoreti, A.~Pereira~do Carmo \emph{et~al.}, ``Multilevel observability in cloud orchestration,'' in \emph{2018 IEEE 16th Intl Conf on Dependable, Autonomic and Secure Computing, 16th Intl Conf on Pervasive Intelligence and Computing, 4th Intl Conf on Big Data Intelligence and Computing and Cyber Science and Technology Congress(DASC/PiCom/DataCom/CyberSciTech)}, 2018, pp. 776--784.

\bibitem{joshi2020integrated}
K.~P. Joshi, L.~Elluri, and A.~Nagar, ``An integrated knowledge graph to automate cloud data compliance,'' \emph{IEEE Access}, vol.~8, pp. 148\,541--148\,555, 2020.

\bibitem{NFR-chung2012non}
L.~Chung, B.~A. Nixon \emph{et~al.}, \emph{Non-functional requirements in software engineering}.\hskip 1em plus 0.5em minus 0.4em\relax Springer Science \& Business Media, 2012, vol.~5.

\bibitem{NFR-paradkar2017}
S.~Paradkar, \emph{Mastering non-functional requirements}.\hskip 1em plus 0.5em minus 0.4em\relax Packt Publishing Ltd, 2017.

\bibitem{NFR-adams2015}
K.~M. Adams \emph{et~al.}, \emph{Nonfunctional requirements in systems analysis and design}.\hskip 1em plus 0.5em minus 0.4em\relax Springer, 2015, vol.~28.

\bibitem{nicoletti2018cross}
L.~Nicoletti, A.~Margheri \emph{et~al.}, ``Cross-cloud management of sensitive data via blockchain: A payslip calculation use case,'' \emph{CEUR Workshop Proceedings}, 2018.

\bibitem{AWSAutoS97:online}
``Aws auto scaling,'' \url{https://aws.amazon.com/autoscaling/}, AWS, 2025.

\bibitem{AzureAut72:online}
``Azure autoscale,'' \url{https://azure.microsoft.com/en-us/features/autoscale/}, Microsoft, 2025.

\bibitem{Autoscal78:online}
``Autoscaling groups of instances  |  compute engine documentation,'' \url{https://cloud.google.com/compute/docs/autoscaler}, Google.

\bibitem{Autoscal15:online}
``Autoscaling,'' \url{https://cloud.ibm.com/docs/cloud-foundry-public?topic=cloud-foundry-public-autoscale_cloud_foundry_apps}, IBM.

\bibitem{henttonen2007integrability}
K.~Henttonen, M.~Matinlassi \emph{et~al.}, ``Integrability and extensibility evaluation from software architectural models--a case study,'' \emph{The Open Software Engineering Journal}, vol.~1, no.~1, pp. 1--20, 2007.

\bibitem{mesbahi2018reliability}
M.~R. Mesbahi, A.~M. Rahmani, and M.~Hosseinzadeh, ``Reliability and high availability in cloud computing environments: a reference roadmap,'' \emph{Human-centric Computing and Information Sciences}, vol.~8, no.~1, pp. 1--31, 2018.

\bibitem{sharma2014framework}
V.~S. Sharma, R.~R. Ramnani, and S.~Sengupta, ``A framework for identifying and analyzing non-functional requirements from text,'' in \emph{4th international workshop on twin peaks of requirements and architecture}, 2014, pp. 1--8.

\bibitem{hoque2018architecture}
S.~Hoque and A.~Miranskyy, ``Architecture for analysis of streaming data,'' in \emph{2018 IEEE Int. Conf. on Cloud Engineering (IC2E)}.\hskip 1em plus 0.5em minus 0.4em\relax IEEE, 2018, pp. 263--269.

\bibitem{hoque2018online}
------, ``Online and offline analysis of streaming data,'' in \emph{2018 IEEE Int. Conf. on Software Architecture Companion (ICSA-C)}.\hskip 1em plus 0.5em minus 0.4em\relax IEEE, 2018, pp. 68--71.

\bibitem{AWSWellA87:online}
``Aws well-architected - build secure, efficient cloud applications,'' \url{https://aws.amazon.com/architecture/well-architected/?wa-lens-whitepapers.sort-by=item.additionalFields.sortDate&wa-lens-whitepapers.sort-order=desc}, AWS, 2024.

\bibitem{IBM-ReferencArchitecture}
``Reference architecture overview | ibm cloud docs,'' \url{https://cloud.ibm.com/docs/framework-financial-services?topic=framework-financial-services-reference-architecture-overview}.

\bibitem{The6Pill41:online}
``The 6 pillars of the aws well-architected framework | aws partner network (apn) blog,'' \url{https://aws.amazon.com/blogs/apn/the-6-pillars-of-the-aws-well-architected-framework/}, 2 2022.

\bibitem{IBMCloud90:online}
``Ibm cloud for financial services,'' \url{https://www.ibm.com/cloud/financial-services}, IBM, 2025.

\bibitem{Bankingi60:online}
``Banking industry architecture: Reference diagram - ibm cloud architecture center,'' \url{https://www.ibm.com/cloud/architecture/architectures/banking/reference-architecture}, IBM, 2025.

\bibitem{CloudNat67:online}
``Cloud native architecture,'' \url{https://architecture.cncf.io/}, 2025.

\bibitem{Enabling61:online}
``Enabling allianz direct's scaling through platform engineering | cloud native architecture,'' \url{https://architecture.cncf.io/architectures/allianz/}, 11 2024.

\bibitem{ScalingA69:online}
``Scaling adobe’s service delivery foundation with a cell-based architecture | cloud native architecture,'' \url{https://architecture.cncf.io/architectures/adobe/}, 11 2024.

\bibitem{goniwada2022cloud}
S.~R. Goniwada, \emph{Cloud Native Architecture and Design: A Handbook for Modern Day Architecture and Design with Enterprise-Grade Examples}.\hskip 1em plus 0.5em minus 0.4em\relax Springer, 2022.

\bibitem{moreno2018orchestrating}
R.~Moreno-Vozmediano, R.~S. Montero \emph{et~al.}, ``Orchestrating the deployment of high availability services on multi-zone and multi-cloud scenarios,'' \emph{Journal of Grid Computing}, vol.~16, no.~1, pp. 39--53, 2018.

\bibitem{fang2011design}
W.~Fang, B.~Jin \emph{et~al.}, ``Design and evaluation of a pub/sub service in the cloud,'' in \emph{2011 Int. Conf. on Cloud and Service Computing}.\hskip 1em plus 0.5em minus 0.4em\relax IEEE, 2011, pp. 32--39.

\bibitem{tran2011eqs}
N.-L. Tran, S.~Skhiri \emph{et~al.}, ``Eqs: An elastic and scalable message queue for the cloud,'' in \emph{2011 IEEE Third Int. Conf. on Cloud Computing Technology and Science}.\hskip 1em plus 0.5em minus 0.4em\relax IEEE, 2011, pp. 391--398.

\bibitem{WhatisPu54:online}
``What is pub/sub? the publish/subscribe model explained,'' \url{https://ably.com/topic/pub-sub}, 5 2023.

\bibitem{Architec53:online}
``Architecting cloud native .net applications for azure,'' \url{https://docs.microsoft.com/en-us/dotnet/architecture/cloud-native/}, Microsoft, 2024.

\bibitem{WhatisCl52:online}
``What is cloud-native? is it hype or the future of software development?'' \url{https://stackify.com/cloud-native/}, 9 2021.

\bibitem{Infrastr92:online}
``Infrastructure as code,'' \url{https://docs.microsoft.com/en-us/dotnet/architecture/cloud-native/infrastructure-as-code}, Microsoft, 6 2022.

\bibitem{RealTime91:online}
``Real-time policy enforcement with governance as code - the new stack,'' \url{https://thenewstack.io/real-time-policy-enforcement-with-governance-as-code/}, 2 2022.

\bibitem{morris2016infrastructure}
K.~Morris, \emph{Infrastructure as code: managing servers in the cloud}.\hskip 1em plus 0.5em minus 0.4em\relax " O'Reilly Media, Inc.", 2016.

\bibitem{GitHubWh52:online}
``Github: Where the world builds software,'' \url{https://github.com/}, GitHub, 2025.

\bibitem{WhatisGi70:online}
``What is gitops?'' \url{https://about.gitlab.com/topics/gitops/}, GitLab, 2025.

\bibitem{GuideToG45:online}
``Guide to gitops,'' \url{https://www.weave.works/technologies/gitops/}, 2024.

\bibitem{beetz2021gitops}
F.~Beetz and S.~Harrer, ``Gitops: The evolution of devops?'' \emph{IEEE Software}, 2021.

\bibitem{GitOpsGi74:online}
``Gitops is continuous deployment for cloud native applications,'' \url{https://www.gitops.tech/}, 2025.

\bibitem{scholl2019cloud}
B.~Scholl, T.~Swanson, and P.~Jausovec, \emph{Cloud Native: Using Containers, Functions, and Data to Build Next-Generation Applications}.\hskip 1em plus 0.5em minus 0.4em\relax " O'Reilly Media, Inc.", 2019.

\bibitem{Schedule90:online}
\BIBentryALTinterwordspacing
(2022) Schedule gpus. Kubernetes. [Online]. Available: \url{https://kubernetes.io/docs/tasks/manage-gpus/scheduling-gpus/}
\BIBentrySTDinterwordspacing

\bibitem{AWSAPIGateway}
``Api management - amazon api gateway - aws,'' \url{https://aws.amazon.com/api-gateway/}, 2025.

\bibitem{AzureAPIGateway}
``Api management – manage api gateways | microsoft azure,'' \url{https://azure.microsoft.com/en-ca/products/api-management}, 2025.

\bibitem{GCPAPIGateway}
``Api gateway  |  google cloud,'' \url{https://cloud.google.com/api-gateway}, 2025.

\bibitem{IBMAPIGateway}
``Api gateway - ibm api connect,'' \url{https://www.ibm.com/products/api-connect/api-gateway}, 2025.

\bibitem{FullyMan11:online}
``Fully managed message queuing – amazon simple queue service,'' \url{https://aws.amazon.com/sqs/}, AWS, 2025.

\bibitem{AzureSer41:online}
``Azure service bus—cloud messaging service,'' \url{https://azure.microsoft.com/en-us/services/service-bus/#overview}, Microsoft, 2025.

\bibitem{PubSubfo88:online}
``Pub/sub for application \& data integration,'' \url{https://cloud.google.com/pubsub}, Google, 2025.

\bibitem{Introduc19:online}
``Introduction to ibm mq,'' \url{https://www.ibm.com/docs/en/ibm-mq/9.0?topic=mq-introduction}, IBM, 9 2021.

\bibitem{birman2007promise}
K.~Birman, ``The promise, and limitations, of gossip protocols,'' \emph{ACM SIGOPS Operating Systems Review}, vol.~41, no.~5, pp. 8--13, 2007.

\bibitem{androutsellis2004survey}
S.~Androutsellis-Theotokis and D.~Spinellis, ``A survey of peer-to-peer content distribution technologies,'' \emph{ACM computing surveys (CSUR)}, vol.~36, no.~4, pp. 335--371, 2004.

\bibitem{pourmajidi2019immutable}
W.~Pourmajidi, L.~Zhang \emph{et~al.}, ``Immutable log storage as a service,'' in \emph{2019 IEEE/ACM 41st Int. Conf. on Software Engineering: Companion Proceedings (ICSE-Companion)}.\hskip 1em plus 0.5em minus 0.4em\relax IEEE, 2019, pp. 280--281.

\bibitem{reilly2010cloud}
D.~Reilly, C.~Wren, and T.~Berry, ``Cloud computing: Forensic challenges for law enforcement,'' in \emph{Int. Conf. for Internet Technology and Secured Transactions (ICITST)}.\hskip 1em plus 0.5em minus 0.4em\relax IEEE, 2010, pp. 1--7.

\bibitem{pourmajidi2018logchain}
W.~Pourmajidi and A.~Miranskyy, ``Logchain: Blockchain-assisted log storage,'' in \emph{2018 IEEE 11th Int. Conf. on Cloud Computing (CLOUD)}.\hskip 1em plus 0.5em minus 0.4em\relax IEEE, 2018, pp. 978--982.

\bibitem{chaer2019blockchain}
A.~Chaer, K.~Salah \emph{et~al.}, ``Blockchain for 5g: Opportunities and challenges,'' in \emph{2019 IEEE Globecom Workshops (GC Wkshps)}.\hskip 1em plus 0.5em minus 0.4em\relax IEEE, 2019, pp. 1--6.

\bibitem{pourmajidi2021immutable}
W.~Pourmajidi, L.~Zhang \emph{et~al.}, ``Immutable log storage as a service on private and public blockchains,'' \emph{IEEE Transactions on Services Computing}, vol.~16, no.~1, pp. 356--369, 2023.

\bibitem{IBM3363o95:online}
``Ibm 3363 optical worm drive,'' \url{https://www.computerhistory.org/collections/catalog/102671161}, Computer History Museum, 11 1987.

\bibitem{WORMAWSS68:online}
``Worm,'' \url{https://aws.amazon.com/blogs/storage/tag/worm/}, AWS.

\bibitem{Overview1:online}
``Overview of immutable storage for blob data - azure storage,'' \url{https://docs.microsoft.com/en-us/azure/storage/blobs/immutable-storage-overview}, Microsoft.

\bibitem{Protecti40:online}
``Protecting cloud storage with worm, key management and more updates,'' \url{https://cloud.google.com/blog/products/storage-data-transfer/protecting-cloud-storage-with-worm-key-management-and-more-updates}, Google.

\bibitem{IBMCloud18:online}
``Ibm cloud object storage - immutable data,'' \url{https://www.ibm.com/ca-en/cloud/object-storage/immutable}, IBM.

\bibitem{obrutsky2016cloud}
S.~Obrutsky, ``Cloud storage: Advantages, disadvantages and enterprise solutions for business,'' in \emph{Conference: EIT New Zealand}, 2016.

\bibitem{AmazonS388:online}
``Amazon s3 - cloud object storage - aws,'' \url{https://aws.amazon.com/s3/}.

\bibitem{FreeClou71:online}
``Free cloud computing services - aws free tier,'' \url{https://aws.amazon.com/free/?all-free-tier}, AWS, 2025.

\bibitem{AmazonS355:online}
``Amazon s3 glacier storage classes,'' \url{https://aws.amazon.com/s3/storage-classes/glacier/}, AWS, 2025.

\bibitem{bass2012software}
L.~Bass, P.~Clements, and R.~Kazman, \emph{Software architecture in practice}, 3rd~ed.\hskip 1em plus 0.5em minus 0.4em\relax Addison-Wesley, 2012.

\bibitem{AWS-RG}
``What are resource groups? - aws resource groups and tags,'' \url{https://docs.aws.amazon.com/ARG/latest/userguide/resource-groups.html}, AWS, 2025.

\bibitem{Azure-RG}
``Manage resource groups - azure portal - azure resource manager,'' \url{https://learn.microsoft.com/en-us/azure/azure-resource-manager/management/manage-resource-groups-portal}, Microsoft, 2025.

\bibitem{Google-RG}
``Using resource groups --- cloud monitoring,'' \url{https://cloud.google.com/monitoring/groups}, Google, 2025.

\bibitem{IBM-RG}
``Managing resource groups,'' \url{https://cloud.ibm.com/docs/account?topic=account-rgs&interface=ui}, IBM, 1 2025.

\bibitem{AmazonKi53:online}
``Amazon kinesis - process \& analyze streaming data,'' \url{https://aws.amazon.com/kinesis/}, AWS, 2025.

\bibitem{EventStr20:online}
``Event streams,'' \url{https://www.ibm.com/cloud/event-streams}, IBM, 2025.

\bibitem{IBMCOS}
``Ibm cloud object storage,'' \url{https://www.ibm.com/products/cloud-object-storage}, 2025.

\bibitem{FastNoSQ6:online}
``Fast nosql key-value database – amazon dynamodb – aws,'' \url{https://aws.amazon.com/dynamodb/}, 2025.

\bibitem{IBMCloudant}
``Ibm cloudant,'' \url{https://www.ibm.com/products/cloudant}.

\bibitem{meehan2017data}
J.~Meehan, C.~Aslantas \emph{et~al.}, ``Data ingestion for the connected world.'' in \emph{Cidr}, vol.~17, 2017, pp. 8--11.

\bibitem{chaudhuri2015governance}
A.~Chaudhuri, ``Governance and risk management in the cloud with cloud controls matrix v3 and iso/iec 38500: 2008,'' in \emph{Handbook of Research on Security Considerations in Cloud Computing}.\hskip 1em plus 0.5em minus 0.4em\relax IGI Global, 2015, pp. 80--101.

\bibitem{saidah2014new}
A.~S. Saidah and N.~Abdelbaki, ``A new cloud computing governance framework.'' in \emph{CLOSER}, 2014, pp. 671--678.

\bibitem{weil2020risk}
T.~Weil, ``Risk assessment methods for cloud computing platforms,'' \emph{IT professional}, vol.~22, no.~1, pp. 63--66, 2020.

\bibitem{ElasticsSearch}
``Elasticsearch: The official distributed search \& analytics engine | elastic,'' \url{https://www.elastic.co/elasticsearch}, 2025.

\bibitem{Logstash}
``Logstash: Collect, parse, transform logs | elastic,'' \url{https://www.elastic.co/logstash}, 2025.

\bibitem{Kibana}
``Kibana: Explore, visualize, discover data | elastic,'' \url{https://www.elastic.co/kibana}, 2025.

\bibitem{Mezmo}
``Telemetry data pipeline \& log analysis solutions | mezmo,'' \url{https://www.mezmo.com/}, 2025.

\bibitem{Splunk}
``Splunk | the key to enterprise resilience,'' \url{https://www.splunk.com/}, 2025.

\bibitem{Sysdig}
``Sysdig | security for containers, kubernetes, and cloud,'' \url{https://sysdig.com/}, 2025.

\bibitem{WilliamPCNA_Governance_Git:online}
``Williampourmajidi/cna\_governance,'' \url{https://github.com/WilliamPourmajidi/cna_governance}, 11 2024.

\bibitem{brunner2015experimental}
S.~Brunner, M.~Bl{\"o}chlinger \emph{et~al.}, ``Experimental evaluation of the cloud-native application design,'' in \emph{2015 IEEE/ACM 8th Int. Conf. on Utility and Cloud Computing (UCC)}.\hskip 1em plus 0.5em minus 0.4em\relax IEEE, 2015, pp. 488--493.

\bibitem{oliveira2017cloud}
F.~Oliveira, S.~Suneja \emph{et~al.}, ``A cloud-native monitoring and analytics framework,'' \emph{IBM Research Division Thomas J. Watson Research Center, Tech. Rep. RC25669 (WAT1710-006)}, 2017.

\bibitem{torkura2017integrating}
K.~A. Torkura, M.~I. Sukmana, and C.~Meinel, ``Integrating continuous security assessments in microservices and cloud native applications,'' in \emph{10th Int. Conf. on Utility and Cloud Computing}, 2017, pp. 171--180.

\bibitem{laszewski2018cloud}
T.~Laszewski, K.~Arora \emph{et~al.}, \emph{Cloud Native Architectures: Design high-availability and cost-effective applications for the cloud}.\hskip 1em plus 0.5em minus 0.4em\relax Packt Publishing Ltd, 2018.

\bibitem{kosinska2020autonomic}
J.~Kosi{\'n}ska and K.~Zieli{\'n}ski, ``Autonomic management framework for cloud-native applications,'' \emph{Journal of Grid Computing}, vol.~18, pp. 779--796, 2020.

\bibitem{toka2021predicting}
L.~Toka, G.~Dobreff \emph{et~al.}, ``Predicting cloud-native application failures based on monitoring data of cloud infrastructure,'' in \emph{2021 IFIP/IEEE Int. Symp. on Integrated Network Management (IM)}.\hskip 1em plus 0.5em minus 0.4em\relax IEEE, 2021, pp. 842--847.

\bibitem{wang2018cloudranger}
P.~Wang, J.~Xu \emph{et~al.}, ``Cloudranger: Root cause identification for cloud native systems,'' in \emph{2018 18th IEEE/ACM Int. Symp. on Cluster, Cloud and Grid Computing (CCGRID)}.\hskip 1em plus 0.5em minus 0.4em\relax IEEE, 2018, pp. 492--502.

\bibitem{DockerAc60:online}
``Docker: Accelerated container application development,'' \url{https://www.docker.com/}.

\bibitem{MessageQ53:online}
``Message queuing service - amazon simple queue service - aws,'' \url{https://aws.amazon.com/sqs/}, 2025.

\bibitem{PushNoti61:online}
``Push notification service - amazon simple notification service - aws,'' \url{https://aws.amazon.com/sns/}.

\bibitem{argocd}
\BIBentryALTinterwordspacing
``Argo cd - declarative gitops cd for kubernetes,'' accessed: 2024-11-09. [Online]. Available: \url{https://argo-cd.readthedocs.io/}
\BIBentrySTDinterwordspacing

\bibitem{fluxcd}
\BIBentryALTinterwordspacing
``Flux cd - the gitops kubernetes operator,'' accessed: 2024-11-09. [Online]. Available: \url{https://fluxcd.io/}
\BIBentrySTDinterwordspacing

\bibitem{jenkinsx}
\BIBentryALTinterwordspacing
``Jenkins x - ci/cd for kubernetes,'' accessed: 2024-11-09. [Online]. Available: \url{https://jenkins-x.io/}
\BIBentrySTDinterwordspacing

\bibitem{spacelift}
\BIBentryALTinterwordspacing
``Spacelift - infrastructure as code management platform,'' accessed: 2024-11-09. [Online]. Available: \url{https://spacelift.io/}
\BIBentrySTDinterwordspacing

\bibitem{DataMode7:online}
``Data model,'' \url{https://zipkin.io/pages/data_model.html}, Zipkin.

\end{thebibliography}

\clearpage
\newpage

\setcounter{page}{1}
\pagenumbering{arabic}

\appendices

\section{Software Engineering and Software Architecture Quality Attribute Alignment}\label{appsec:SoftwareEngineeringQualityAttributes}

Ensuring that reference architectures align with established software engineering and architecture quality attributes is critical for their effectiveness in system design and operational excellence. These attributes act as fundamental benchmarks to validate the architecture's capability to meet functional and non-functional requirements, including aspects like maintainability, scalability, and security.

Table~\ref{table:software_engineering_quality_attributes} provides a detailed alignment of our proposed reference architecture with these quality attributes. The table categorizes the alignment into two levels:
\begin{itemize}
    \item \textbf{Fully Supported:} Indicates direct alignment through the architecture's design or the flexibility to integrate solutions seamlessly.
    \item \textbf{Partially Supported:} Reflects cases where the architecture meets many but not all requirements, requiring additional adopter intervention.
\end{itemize}

This analysis reveals that the proposed architecture fully supports 86\% (12 out of 14) of the attributes, demonstrating its robustness in meeting critical governance needs for CNAs. The remaining 14\% (2 out of 14) are partially supported, emphasizing areas where customization or additional solutions may be necessary. For further discussion of the role these attributes play in software engineering and architecture design, readers are encouraged to refer to Section~\ref{subsubsec:Software Engineering and Software Architecture Comparison}.

\begin{table*}[bt]

\caption{Software Engineering and Software Architecture Quality Attribute Alignment.}
\label{table:software_engineering_quality_attributes}
\resizebox{\textwidth}{!}{

\begin{tabular}{@{}p{0.20\textwidth}|p{0.4\textwidth}|p{0.4\textwidth}@{}}
\toprule
\textbf{Quality Attribute} & \textbf{Description} & \textbf{Our Architecture Alignment} \\ \midrule
Availability & Focuses on the system's readiness for operation, encompassing reliability and self-repair capabilities. & Fully Supported: Implements redundancy, monitoring, and failover mechanisms. \\ \midrule
Interoperability & Ability of systems to exchange and interpret information correctly through interfaces. & Fully Supported: Utilizes standardized interfaces for seamless external interactions. \\ \midrule
Modifiability & Ease of system modifications for functionalities, fault corrections, and performance improvements. & Partially Supported: Highly modular design, though limited by some components. \\ \midrule
Performance & System's capability to meet timing requirements and efficiently process events. & Fully Supported: Optimizes for performance with scalable solutions. \\ \midrule
Security & Protection against unauthorized access or modifications to ensure data confidentiality, integrity, and availability. & Fully Supported: Integrates comprehensive security measures. \\ \midrule
Testability & Ease with which the software can be tested to confirm it meets its requirements. & Fully Supported: Facilitates testing with automated frameworks and CI processes. \\ \midrule
Usability & User-friendliness and efficiency in accomplishing desired tasks with the system. & Fully Supported: Emphasizes intuitive interfaces and user support. \\ \midrule
Portability & Ease of transferring software across different platforms or environments. & Fully Supported: Ensures portability with cloud-native principles. \\ \midrule
Development Distributability & Facilitation of distributed software development processes. & Fully Supported: Employs tools supporting global team collaboration. \\ \midrule
Scalability & Ability to handle increased loads by scaling resources. & Fully Supported: Elastic resources and load balancing ensure dynamic demand management. \\ \midrule
Deployability & Ease of deploying software across environments. & Fully Supported: Automated pipelines and tools streamline deployments. \\ \midrule
Mobility & Software's adaptability to different platforms or devices, maintaining user experience. & Partially Supported: Responsive design for platform adaptability, with some complex application limits. \\ \midrule
Monitorability & Capability for real-time system operation monitoring. & Fully Supported: Integrates monitoring tools for performance and health insights. \\ \midrule
Safety & Software's capability to avoid causing harm or loss. & Fully Supported: Emphasizes error handling and safety standards compliance. \\ \bottomrule
\end{tabular}
}
\end{table*}

\section{Alignment of Proposed Architecture with Cloud Controls Matrix Domains}\label{appsec:CCMAlignment}

The Cloud Controls Matrix (CCM), developed by the Cloud Security Alliance (CSA), provides a comprehensive framework for assessing the security, compliance, and governance of cloud services across 17 domains. These domains cover critical areas such as data security, identity management, and business continuity, offering a robust benchmark for cloud governance.

Table~\ref{table:CCM_alignment} evaluates the alignment of our proposed reference architecture with the CCM domains. The table categorizes the architecture's support as follows:
\begin{itemize}
    \item \textbf{Fully Supported:} Indicates that the architecture directly addresses the domain requirements, embedding the necessary features and practices.
    \item \textbf{Partially Supported:} Reflects cases where the architecture meets several but not all requirements, necessitating additional efforts from adopters.
    \item \textbf{Out of Scope:} Domains that are not within the primary focus of the proposed architecture, such as human resource security or mobile-specific concerns.
\end{itemize}

The alignment analysis reveals the following:
\begin{itemize}
    \item \textbf{11 domains (65\%)} are fully supported, showcasing robust capabilities in areas like audit and assurance, encryption, and threat management.
    \item \textbf{4 domains (24\%)} are partially supported, highlighting areas such as business continuity and supply chain management where additional customization may be required.
    \item \textbf{2 domains (12\%)} are out of scope, reflecting the architecture's emphasis on technological governance rather than organizational or mobile-specific aspects.
\end{itemize}

This evaluation demonstrates the architecture's strong alignment with CCM domains, ensuring comprehensive support for cloud governance while recognizing areas for potential adopter intervention. For additional discussion on these domains, refer to Section~\ref{subsec:Cloud Control Matrix Comparison}.

\begin{table*}[bt]
\caption{Alignment of Proposed Architecture with Cloud Controls Matrix Domains.}
\label{table:CCM_alignment}
\resizebox{\textwidth}{!}{

\begin{tabular}{@{}p{0.3\textwidth}|p{0.35\textwidth}|p{0.35\textwidth}@{}}
\toprule
\textbf{CCM Domain} & \textbf{Summary} & \textbf{Alignment \& Rationale} \\ \midrule
Audit \& Assurance & Ensures proper audit trails and assurance processes are in place. & Fully Supported: Our architecture automates audit logs and integrates with assurance standards. \\ \midrule
Application \& Interface Security & Secures applications and their interfaces against unauthorized access and attacks. & Fully Supported: Secure coding and interface protocols are core to our design. \\ \midrule
Business Continuity Management \& Operational Resilience & Maintains operations through disruptions, ensuring business continuity. & Partially Supported: Includes basic resilience features; advanced strategies need customization. \\ \midrule
Change Control \& Configuration Management & Manages changes to systems and configurations to maintain security. & Fully Supported: Automated change control and configuration management are embedded. \\ \midrule
Data Security \& Information Lifecycle Management & Protects data throughout its lifecycle, from creation to deletion. & Fully Supported: Implements strict data handling, encryption, and deletion policies. \\ \midrule
Datacenter Security & Focuses on securing the physical and virtual infrastructure of data centers. & Partially Supported: Virtual security is robust; physical aspects depend on deployment. \\ \midrule
Encryption \& Key Management & Manages encryption practices and key lifecycle. & Fully Supported: Integrates advanced encryption and key management standards. \\ \midrule
Governance, Risk \& Compliance & Aligns IT with business objectives while managing risk and meeting compliance. & Fully Supported: Inbuilt GRC framework to comply with legal and regulatory standards. \\ \midrule
Human Resources Security & Addresses security measures related to hiring, training, and managing personnel. & Out of Scope: Focuses on technological aspects of security, not HR practices. \\ \midrule
Identity \& Access Management& Manages users' identities and their access to resources. & Fully Supported: Advanced IAM solutions with extensive access controls and authentication methods. \\ \midrule
Infrastructure \& Virtualization Security & Secures the infrastructure, especially in virtualized environments. & Fully Supported: Security isolation, access control, and monitoring for virtual resources. \\ \midrule
Interoperability \& Portability & Ensures systems work across different environments and can be moved easily. & Fully Supported: Standards-based APIs and container technologies ensure flexibility. \\ \midrule
Mobile Security & Protects against threats specific to mobile platforms and devices. & Out of Scope: Architecture targets broader cloud environments, not mobile-specific issues. \\ \midrule
Security Incident Management, E-Discovery \& Cloud Forensics & Manages security incidents with procedures for discovery and forensic analysis. & Fully Supported: Robust incident response, e-discovery, and forensic tools integrated. \\ \midrule
Supply Chain Management, Transparency \& Accountability & Ensures security and transparency throughout the supply chain. & Partially Supported: Mechanisms for software supply chain visibility; enhanced with third-party tools. \\ \midrule
Threat \& Vulnerability Management & Identifies and mitigates vulnerabilities and manages threats. & Fully Supported: Proactive threat identification and mitigation embedded in the architecture. \\ \midrule
Universal Endpoint Management& Manages and secures endpoints within the cloud environment. & Partially Supported: Basic management included; extended functionality via third-party integrations. \\ \bottomrule
\end{tabular}
}
\end{table*}

\section{Real-world Instantiation of the Proposed Reference Architecture}\label{appsec:Real-world Instantiation of the Proposed Reference Architecture}

In this section, we present a real-world implementation of the proposed reference architecture by selecting and deploying a subset of its components. As illustrated in Figure~\ref{fig:dataflow}, the architecture is divided into three distinct RGs: \RG{RG-1}, \RG{RG-GOV-IMS}, and \RG{RG-GOV-DA}. Section~\ref{subsec:Subset Selection} provides the rationale for selecting a subset of the proposed reference architecture components to be implemented. Section~\ref{subsec:CNA Implementation} provides details on the implementation of the CNAs that will generate and emit telemetry and logs to the proposed reference architecture. Section~\ref{subsec:AWS-RG-1: High-Demand CNA Environment} describes a high-demand CNA environment implemented on AWS cloud by employing scalable components of our governance architecture. Section~\ref{subsec:IBM-RG-1: Less-Demanding CNA Environment} demonstrates the architecture's flexibility by directly forwarding telemetry from a less-demanding CNA, implemented on the IBM Cloud, to the governance pipeline. Section~\ref{subsec:RG-GOV-IMS: Ingestion, Manipulation, and Storage} describes the implementation of \RG{RG-GOV-IMS} as the primary processing hub responsible for telemetry ingestion, manipulation, and storage. Section~\ref{subsec:RG-GOV-DA: Data Analytics} explains the implementation of \RG{RG-GOV-DA} as a data analytics engine to analyze telemetry stored in immutable and mutable storage options. Section~\ref{subsec:Results} explores the outcomes of implementing the proposed governance pipeline for AWS and IBM Cloud CNAs. The results provide insights into the system's performance, highlighting differences in processing times across various stages.

The GitHub repository for this implementation, encompassing a subset of the CNA Governance Platform reference architecture, is described in detail in Appendix~\ref{appsubsec:Git Repository}. The repository includes sample CNA code and Terraform configurations for deploying selected components across multi-cloud environments on AWS and IBM. It adheres to GitOps principles, organizing infrastructure as code into modular components within distinct resource groups.

\subsection{Subset Selection}\label{subsec:Subset Selection}

\begin{figure}[ht]
\centering
\includegraphics[width=\columnwidth]{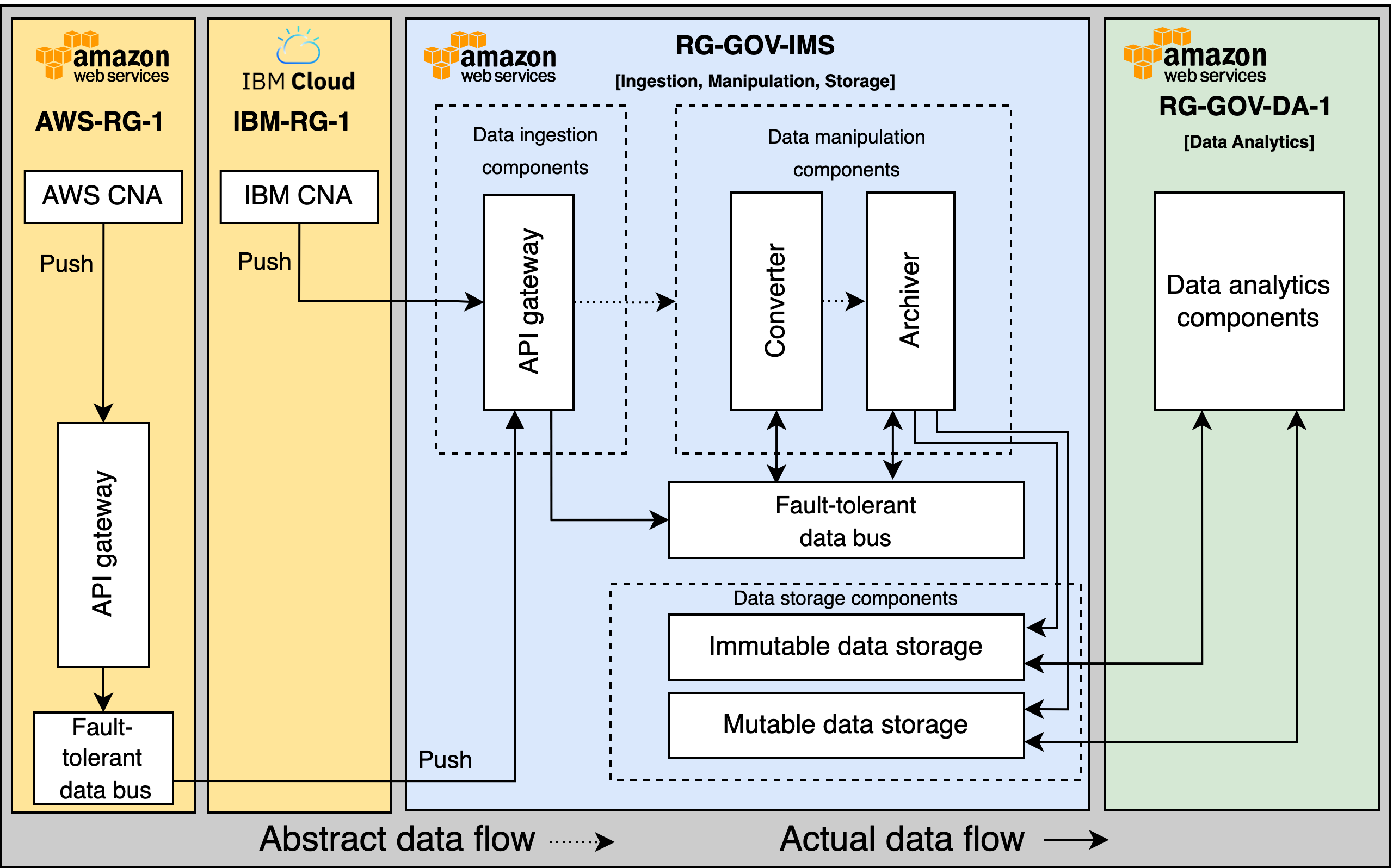}
\caption{Graphical representation of the implemented subset of the reference architecture.}
\label{fig:implemented-subset}
\end{figure}

The selection of components for implementing the subset of the proposed reference architecture was guided by two key principles: showcasing a representative subset of governance components and ensuring that these components are practical examples that practitioners can extend upon. Figure~\ref{fig:implemented-subset} depicts the selected subset. To make the instantiation demonstrating a multi-cloud deployment,~\RG{RG-1} is hosted on AWS, and~\RG{RG-2} is hosted on IBM. The proposed reference architecture, presented in ~\RG{RG-GOV-IMS} and ~\RG{RG-GOV-DA} are hosted on AWS.

\subsection{CNA Implementation}\label{subsec:CNA Implementation}

To simulate a real-world CNA scenario, we developed a CNA using Python. This CNA is containerized using Docker~\cite{DockerAc60:online}, allowing it to be deployed across multiple cloud platforms. The CNA continuously generates telemetry data. For the purposes of this experiment, the telemetry data consists of metrics and logs. CNA emits this telemetry every second, simulating a high-demand application environment.

In this experiment, we ran the CNAs for 4,000 seconds, resulting in the generation of 4,000 telemetry entries, comprising 2,000 metrics and 2,000 logs. The metrics are stored in mutable storage, reflecting their less critical role, while logs are archived in immutable storage to ensure the tamper-proof requirements necessary for digital evidence. AWS and IBM instances of the app independently generated 1,000 metrics and 1,000 logs, contributing equally to the total telemetry data set of 4,000 entries. This data set is available in the project repository~\cite{WilliamPCNA_Governance_Git:online}.

As part of the multi-cloud deployment scenario, we implemented two CNA instances:

\begin{itemize}
    \item \textbf{AWS CNA.} Deployed on AWS cloud in \RG{AWS-RG-1}.
    \item \textbf{IBM CNA.} Deployed on IBM Cloud in \RG{IBM-RG-1}.
\end{itemize}
This design simulates the telemetry flow of the same CNA deployed across multiple clouds, providing redundancy and scalability. Both CNA instances share the same architecture and characteristics and generate two types of data:
    \begin{itemize}
        \item \textbf{Telemetry Data.} Includes metrics such as CPU, memory, disk, and network utilization.
        \item \textbf{Application Logs.} Includes operational logs that describe the internal processes of the CNA.
    \end{itemize}
More information on CNA telemetry, including the format and samples of JSON files, is provided in Appendix~\ref{appsec:JSON Structure and Tags}.

\subsection{AWS-RG-1: High-Demand CNA Environment}\label{subsec:AWS-RG-1: High-Demand CNA Environment}

In \RG{AWS-RG-1}, we simulated a scenario where the CNA is deployed in a high-demand environment, continuously generating a large volume of telemetry data. Due to the high velocity of data, additional components were introduced to handle the incoming telemetry effectively, ensuring fault tolerance and scalability.

In addition to the CNA, two key components were implemented in ~\RG{AWS-RG-1} as follows.

\subsubsection{AWS API Gateway} The telemetry data generated by the AWS CNA is first received by an API Gateway. The API Gateway acts as the entry point for the governance pipeline in~\RG{AWS-RG-1}, allowing telemetry to be ingested in a scalable and secure manner. The API Gateway is configured to trigger a Lambda function whenever telemetry data are received.
    
\subsubsection{AWS Lambda Function} The AWS Lambda function processes the telemetry data, appending the timestamp at which the telemetry was received by the API Gateway. This is represented by the tag \texttt{RG\_1\_API\_Gateway\_timestamp}. Once the telemetry data is processed, it is forwarded to a fault-tolerant data bus.
    
\subsubsection{AWS Simple Queue Service (SQS)} As the fault-tolerant data bus, AWS SQS~\cite{MessageQ53:online} is employed to temporarily store the telemetry data before they continue through the governance pipeline. The Lambda function sends the telemetry data to AWS SQS, ensuring that the high-velocity data are managed efficiently. AWS SQS also ensures durability by providing fault tolerance in case of failures. A second Lambda function attached to the SQS queue processes the stored telemetry, adding the tag \texttt{RG\_1\_SQS\_Forwarder\_timestamp}, and submits it to the next stage of the pipeline in \RG{RG-GOV-IMS}. Once the telemetry data reaches \RG{RG-GOV-IMS}, both the AWS and IBM CNA telemetry flow through the same governance pipeline.

The implementation in \RG{AWS-RG-1} reflects the handling of a demanding CNA that requires a robust pipeline for processing high-velocity data and ensuring fault tolerance and scalability.

\subsection{IBM-RG-1: Less-Demanding CNA Environment}\label{subsec:IBM-RG-1: Less-Demanding CNA Environment}

In \RG{IBM-RG-1}, the CNA operates in a less-demanding environment. Unlike AWS, the IBM CNA submits telemetry directly to the \component{API Gateway} in \RG{RG-GOV-IMS} without requiring an API Gateway or a dedicated data bus within the same RG. This implementation showcases the flexibility of the governance architecture to handle different levels of demand, streamlining the process for environments with lower telemetry needs.

\subsection{RG-GOV-IMS: Ingestion, Manipulation, and Storage}\label{subsec:RG-GOV-IMS: Ingestion, Manipulation, and Storage}

\RG{RG-GOV-IMS} serves as the primary resource group in the proposed reference architecture, focusing on data ingestion, manipulation, and storage. This resource group implements core governance components that facilitate the processing of telemetry and logs from cloud-native applications deployed across multiple clouds. The architecture is designed to support scalability, flexibility, and cross-cloud data governance.

The~\RG{RG-GOV-IMS} is structured with three major components: Data Ingestion, responsible for receiving and routing incoming telemetry; Data Manipulation, which processes the telemetry before archiving; and Data Storage, offering both mutable and immutable storage options based on governance policies.

Given the need for topic-based messaging to support multiple subscribers, we opted to use AWS Simple Notification Service (SNS) instead of AWS SQS. AWS SNS is a fully managed pub/sub messaging service that supports FIFO messaging~\cite{PushNoti61:online}, ensuring the order and reliability of message delivery. Additionally, AWS API Gateway is employed to receive telemetry data, while Lambda functions act as serverless computing components for processing the data.

\subsubsection{Data Ingestion}
In the \RG{RG-GOV-IMS} resource group, the AWS API Gateway serves as the entry point for telemetry data coming from both \RG{AWS-RG-1} and \RG{IBM-RG-1}. The API Gateway receives telemetry submissions and immediately triggers a Lambda function. This function processes the telemetry by adding a timestamp (\texttt{RG\_GOV\_IMS\_API\_Gateway\_timestamp}) and forwarding the message to the SNS topic \texttt{converter}. This step represents the first phase in the ingestion process, where telemetry is collected and routed for further manipulation.

\subsubsection{Data Manipulation}
The data manipulation phase begins when the SNS topic ``converter'' is triggered. A Lambda function, referred to as the \texttt{Converter}, is subscribed to the ``converter'' topic and processes the incoming telemetry. The converter adds a new timestamp (\texttt{RG\_GOV\_IMS\_Converter\_timestamp}) and forwards the data to the same SNS topic under the \texttt{Archiver} module.

\subsubsection{Data Storage}
The \texttt{Archiver}, another Lambda function, is responsible for the final manipulation step. It receives telemetry from the \texttt{Archiver} topic, adds its own timestamp (\texttt{RG\_GOV\_IMS\_Archiver\_timestamp}), and proceeds to prepare the data for storage. The module determines the appropriate storage destination based on the \texttt{data\_type} tag within the telemetry. It differentiates between metrics, logs, and traces, storing each type in the appropriate bucket as follows.
\paragraph{Mutable S3 Storage} If the \texttt{data\_type} is \texttt{metrics}, the telemetry is stored in a mutable S3 bucket without object lock.
\paragraph{Immutable S3 Storage} If the \texttt{data\_type} is \texttt{logs} or \texttt{traces}, the telemetry is stored in an immutable S3 bucket with object lock enabled. This ensures the data is protected from modification or deletion for a predefined retention period of one year, adhering to strict governance requirements.

During this process, the archiver generates a unique object key for each telemetry message based on the \texttt{CSP} and \texttt{log\_id}. This key is used to name the object when storing it in the S3 bucket, ensuring consistent traceability across the governance pipeline.

By integrating timestamps at every stage of the data flow, \RG{RG-GOV-IMS} enables comprehensive time-delay analysis between components, allowing practitioners to measure the performance of the governance pipeline. This structured ingestion, manipulation, and storage process ensures optimal handling of telemetry data in a multi-cloud governance environment.

\subsection{RG-GOV-DA: Data Analytics}\label{subsec:RG-GOV-DA: Data Analytics}

\RG{RG-GOV-DA} serves as a dedicated RG for various types of governance-related data analytics. This RG is designed to support a wide range of analytics engines, such as anomaly detection, drift detection, SLA breach detection, and compliance breach detection. Its modular structure provides flexibility, allowing practitioners to deploy specific analytics modules based on their organizational needs. Multiple instances of \RG{RG-GOV-DA} can be implemented to serve different departments or teams, depending on the scope of their governance analytics requirements.

For the purposes of this experiment, we focused on a data analytics engine that computes time delays between key components in the governance pipeline, with telemetry data stored in immutable storage. The choice of immutable storage, such as S3 with object lock, ensures that the governance data is tamper-proof, which is essential in regulatory and compliance scenarios where data integrity is critical. The engine specifically analyzes data from immutable storage, as this is a more realistic scenario in governance frameworks, where data must remain unaltered for audit and compliance purposes.

The engine calculates the delays, or time responses, between components in the CNAs deployed on AWS and IBM Cloud. This analysis simulates real-world use cases where Service Level Agreements (SLAs) impose strict time constraints on system response times, making it crucial to monitor and analyze the performance of each component in the pipeline.

\subsubsection{Data Ingestion and Serverless Analytics}
The time response data analytics engine is implemented using AWS Lambda functions, offering a serverless solution that dynamically scales based on the volume of data processed. The engine is invoked on demand, retrieving telemetry data from S3 buckets, calculating time delays, and generating visual reports, including box plots. The serverless nature ensures that the engine is cost-effective and scalable, handling large datasets efficiently without requiring a permanent infrastructure.

\subsubsection{Leg Definition and Results}
We analyzed time delays across the following legs for AWS-based and IBM-cloud-based CNAs:
\begin{itemize}
    \item \textbf{AWS Leg 1} $=$ RG\_1\_API\_Gateway\_timestamp~$-$ cna\_timestamp,
    \item \textbf{AWS Leg 2} $=$ RG\_1\_SQS\_Forwarder\_timestamp~$-$ RG\_1\_API\_Gateway\_timestamp,
    \item \textbf{AWS Leg 3} $=$ RG\_GOV\_IMS\_API\_Gateway\_timestamp~$-$ RG\_1\_SQS\_Forwarder\_timestamp,
    \item \textbf{AWS Leg 4} $=$ RG\_GOV\_IMS\_Converter\_timestamp~$-$ RG\_GOV\_IMS\_API\_Gateway\_timestamp,
    \item \textbf{AWS Leg 5} $=$ RG\_GOV\_IMS\_Archiver\_timestamp~$-$ RG\_GOV\_IMS\_Converter\_timestamp,
    \item \textbf{IBM Leg 1} $=$ RG\_GOV\_IMS\_API\_Gateway\_timestamp~$-$ cna\_timestamp,
    \item \textbf{IBM Leg 4} $=$ RG\_GOV\_IMS\_Converter\_timestamp~$-$ RG\_GOV\_IMS\_API\_Gateway\_timestamp,
    \item \textbf{IBM Leg 5} $=$ RG\_GOV\_IMS\_Archiver\_timestamp~$-$ RG\_GOV\_IMS\_Converter\_timestamp,
\end{itemize}
where cna\_timestamp represents the timestamps when telemetry data leaves an instance of our CNA app.

\subsection{Results}\label{subsec:Results}
\begin{figure*}[tb]
    \centering
    \includegraphics[width=\textwidth]{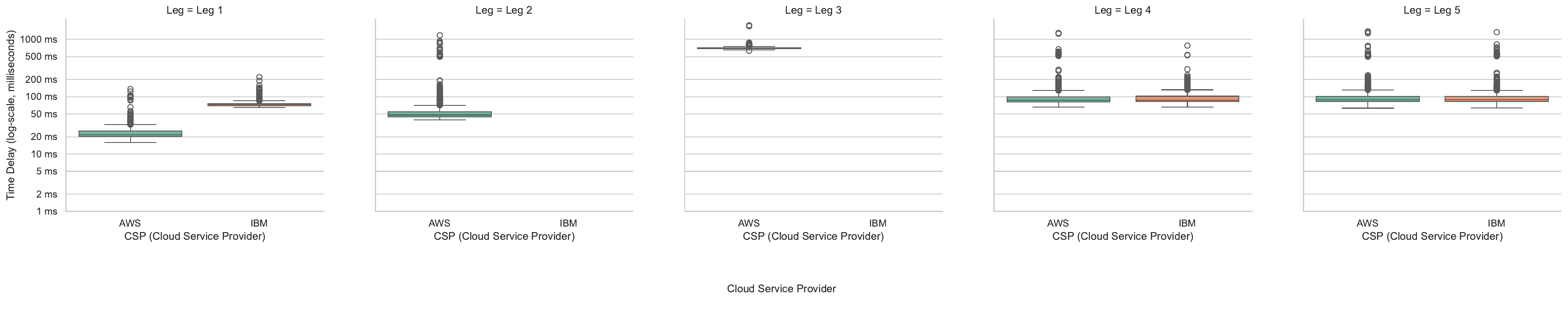}
    \caption{Box Plot: Log-Scaled Time Delay Comparison Across Legs for AWS and IBM Cloud.}
    \label{fig:stats}
\end{figure*}

Table~\ref{tbl:stats} provides information on the average and variability of the delays in both AWS and IBM Cloud CNAs, enabling a comparison between the two providers. Figure~\ref{fig:stats} presents a comparative box plot illustrating the time delays across each leg for both AWS and IBM Cloud. The analysis of the time delays data in Table~\ref{tbl:stats}, and Figure~\ref{fig:stats} shows interesting variations across the legs for AWS and IBM Cloud.

For Leg 1, AWS shows lower delays compared to IBM Cloud, which is expected due to the geographical proximity of AWS CNA and API Gateway within the same region (us-east-1). IBM Cloud experiences higher delays in Leg 1 due to the need to transmit data between different regions (ca-tor to us-east-1). The higher delays observed in IBM Cloud for Leg 1 are likely due to this inter-regional data transmission, where cross-cloud communication adds latency.

For Leg 4 and Leg 5, the delays for both AWS and IBM Cloud are comparable, as they share the same governance pipeline. This consistency in performance between the two CSPs indicates that once the data reaches the governance pipeline, the time delays are largely influenced by the pipeline itself rather than the originating CSP. The comparable values highlight consistency in processing times across both cloud providers for these stages.

AWS-specific Legs 2 and 3 show delays unique to AWS's internal processing, where messages are forwarded through SQS and processed by the RG\_GOV\_IMS\_API\_Gateway. These legs reflect the time spent forwarding messages through the SQS system and invoking the RG\_GOV\_IMS\_API\_Gateway.IBM Cloud does not have corresponding legs, as its CNA directly interacts with the RG\_GOV\_IMS\_API\_Gateway, bypassing these additional steps.

\begin{table}[tb]
    \centering
    \caption{Summary statistics~---~including the minimum, maximum, mean, median, and standard deviation (std)~---~of time delays across different legs for AWS and IBM Cloud. }
    \label{tbl:stats}
    \begin{tabularx}{\linewidth}{@{}X|X|>{\raggedleft\arraybackslash}X|>{\raggedleft\arraybackslash}X|>{\raggedleft\arraybackslash}X|>{\raggedleft\arraybackslash}X|>{\raggedleft\arraybackslash}X@{}}
        \toprule
        \textbf{Leg} & \textbf{CSP} & \textbf{Min (ms)} & \textbf{Max (ms)} & \textbf{Mean (ms)} & \textbf{Median (ms)} & \textbf{Std (ms)} \\
        \midrule
        Leg 1 & AWS & 15.78 & 136.03 & 23.61 & 21.69 & 7.08 \\
        Leg 1 & IBM & 64.40 & 219.08 & 74.83 & 72.43 & 11.38 \\
        Leg 2 & AWS & 39.42 & 1173.62 & 63.27 & 47.89 & 79.44 \\
        Leg 3 & AWS & 630.42 & 822.80 & 705.04 & 697.59 & 29.77 \\
        Leg 4 & AWS & 66.14 & 611.94 & 103.56 & 85.89 & 68.27 \\
        Leg 4 & IBM & 65.57 & 778.71 & 98.48 & 87.18 & 38.99 \\
        Leg 5 & AWS & 63.11 & 737.70 & 109.59 & 87.65 & 80.96 \\
        Leg 5 & IBM & 63.27 & 1332.78 & 109.65 & 87.43 & 90.92 \\
        \bottomrule
    \end{tabularx}
\end{table}

\subsection{Git Repository}\label{appsubsec:Git Repository}

This repository implements a subset of the CNA Governance Platform reference architecture, designed to manage, process, and archive telemetry data generated by CNAs deployed across multi-cloud environments on AWS and IBM. The full implementation, including deployment instructions, usage details, and explanations for each component, is available on GitHub~\cite{WilliamPCNA_Governance_Git:online}.

The repository adheres to GitOps principles, organizing infrastructure as code into modular components within distinct RGs:

\begin{itemize}
    \item \texttt{cna/}: Contains the CNAs developed for AWS and IBM, responsible for generating telemetry and logs. Each CNA is containerized using Docker and includes separate \texttt{README.md} files with deployment instructions for each cloud.
    \item \texttt{sample\_cna\_generated\_data/}: Provides sample telemetry (metrics and logs) generated by the AWS and IBM CNAs. This data demonstrates the telemetry format and structure processed by the platform, organized under \texttt{aws/} and \texttt{ibm/} folders.
    \item \texttt{terraform/}: Includes Terraform configurations for deploying AWS infrastructure, divided into two main resource groups:
        \begin{itemize}
            \item \texttt{AWS-RG-1}: Contains configurations for API Gateway, SQS, and Lambda functions for initial telemetry ingestion and forwarding.
            \item \texttt{AWS-RG-GOV-IMS}: Contains configurations for the API Gateway, SNS topics (Converter and Archiver), Lambda functions, and S3 storage (mutable and immutable), handling further telemetry data processing, conversion, and archiving.
        \end{itemize}
    \item \texttt{AWS-RG-GOV-DA/}: Contains a sample analytics application that processes archived telemetry data from S3 storage. The application calculates processing delays across telemetry processing stages, providing statistical analysis and visualization.
\end{itemize}

Each folder and subfolder in the repository contains a \texttt{README.md} file with instructions on deployment, usage, and a description of each component's purpose.

\subsection{GitOps Considerations}

To effectively implement the CNA Governance Platform reference architecture, we recommend adopting GitOps principles to automate and streamline application delivery. By treating IaC and utilizing containerization, as demonstrated through our provided Terraform configurations and Docker setups, teams can achieve consistent and reproducible deployments. Leveraging CI/CD pipelines, along with DevOps and DevSecOps tools, enables single-click deployments, enhancing efficiency and reducing manual errors. Integrating security practices within these pipelines ensures that security is an integral part of the deployment process, not an afterthought. By embracing these methodologies, organizations can maintain a robust, secure, and agile deployment workflow, aligning with modern best practices in software development and operations. To facilitate the adoption of GitOps principles, one can apply the same principles using other commonly used GitOps tools such as Argo CD~\cite{argocd}, Flux CD~\cite{fluxcd}, Jenkins X~\cite{jenkinsx}, and Spacelift~\cite{spacelift}.

\section{JSON Structure and Tags}\label{appsec:JSON Structure and Tags} The CNA submits both metrics and logs in a JSON format inspired by the Zipkin~\cite{DataMode7:online} tracing format. The structure includes several key elements, which we describe below, followed by examples of AWS and IBM metrics and logs.

\begin{itemize}
    \item \textbf{CSP Tag.} Identifies the Cloud Service Provider hosting the CNA. The value is either \texttt{"IBM"} or \texttt{"AWS"}.

    \item \textbf{Data Type Tag.} Distinguishes between telemetry and log data using the \texttt{"data\_type"} field, which can take the values \texttt{"telemetry"} or \texttt{"logs"}. Examples of AWS JSON data for these types are provided in Figures~\ref{lst:json-aws-metrics} and~\ref{lst:json-aws-logs}. Examples of IBM JSON data are provided in Figures~\ref{lst:json-ibm-metrics} and~\ref{lst:json-ibm-logs}.

    \item \textbf{Timestamps Tag.} Contains a nested JSON structure that tracks the generation time of telemetry and logs. Each governance component appends its own timestamp to the \texttt{"timestamps"} field, allowing time-delay analysis between components.

    \item \textbf{Unique ID.} Each telemetry or log message is assigned a unique identifier, ensuring traceability throughout the governance pipeline. The unique ID is also used by the \component{archiver} component for naming objects stored in either mutable or immutable storage.
\end{itemize}

\begin{figure*}
\begin{lstlisting}[language=json]
{
  "CSP": "AWS",
  "data_type": "metrics",
  "error": null,
  "governance_data": {
    "memory_usage": 37.6,
    "cpu_usage": 1.5,
    "disk_usage": 34.3,
    "bytes_sent": 8375883,
    "bytes_recv": 82911423,
    "additional_metric_1": "value_1",
    "additional_metric_2": "value_2"
  },
  "log_id": "a2bc98fb-d4f2-44b0-a093-3f766fd1016e",
  "service_name": "cna-app",
  "timestamps": {
    "cna_timestamp": "2024-09-06T15:13:45.460268+00:00",
    "RG_1_API_Gateway_timestamp": "2024-09-06T15:13:45.478742+00:00",
    "RG_1_SQS_Forwarder_timestamp": "2024-09-06T15:13:45.524243+00:00",
    "RG_GOV_IMS_API_Gateway_timestamp": "2024-09-06T15:13:46.232296+00:00",
    "RG_GOV_IMS_Converter_timestamp": "2024-09-06T15:13:46.333993+00:00",
    "RG_GOV_IMS_Archiver_timestamp": "2024-09-06T15:13:46.444789+00:00"
  }
}
\end{lstlisting}
\caption{Sample AWS JSON with data type metrics}
\label{lst:json-aws-metrics}
\end{figure*}

\begin{figure*}
\begin{lstlisting}[language=json]
{
  "CSP": "IBM",
  "data_type": "metrics",
  "error": null,
  "governance_data": {
    "memory_usage": 36.5,
    "cpu_usage": 2.0,
    "disk_usage": 4.2,
    "bytes_sent": 1770526,
    "bytes_recv": 4713567,
    "additional_metric_1": "value_1",
    "additional_metric_2": "value_2"
  },
  "log_id": "0cc2de3f-dd47-40a9-86fe-62c1a1ca0c22",
  "service_name": "cna-app",
  "timestamps": {
    "cna_timestamp": "2024-08-12T17:21:47.260626+00:00",
    "RG_GOV_IMS_API_Gateway_timestamp": "2024-08-12T17:21:47.327862+00:00",
    "RG_GOV_IMS_Converter_timestamp": "2024-08-12T17:21:47.422773+00:00",
    "RG_GOV_IMS_Archiver_timestamp": "2024-08-12T17:21:47.557046+00:00"
  }
}
\end{lstlisting}
\caption{Sample IBM JSON with data type metrics}
\label{lst:json-ibm-metrics}
\end{figure*}

\begin{figure*}
\begin{lstlisting}[language=json]
{
  "CSP": "AWS",
  "data_type": "logs",
  "error": null,
  "governance_data": {
    "log_1": "Application started successfully.",
    "log_2": "Collecting system metrics.",
    "log_3": "Metrics collected successfully.",
    "log_4": "Sending data to API Gateway.",
    "log_5": "Data sent successfully.",
    "log_6": "Error handling and logging mechanism operational.",
    "log_7": "System monitoring and logging active.",
    "log_8": "Routine check completed successfully.",
    "log_9": "No errors detected in the last cycle.",
    "log_10": "All systems functional."
  },
  "log_id": "a18f5984-3ead-43c7-af73-05f3cf5bce6f",
  "service_name": "cna-app",
  "timestamps": {
    "cna_timestamp": "2024-09-06T14:48:46.268925+00:00",
    "RG_1_API_Gateway_timestamp": "2024-09-06T14:48:46.290639+00:00",
    "RG_1_SQS_Forwarder_timestamp": "2024-09-06T14:48:46.332218+00:00",
    "RG_GOV_IMS_API_Gateway_timestamp": "2024-09-06T14:48:47.051179+00:00",
    "RG_GOV_IMS_Converter_timestamp": "2024-09-06T14:48:47.141164+00:00",
    "RG_GOV_IMS_Archiver_timestamp": "2024-09-06T14:48:47.222075+00:00"
  }
}
\end{lstlisting}
\caption{Sample AWS JSON with data type logs}
\label{lst:json-aws-logs}
\end{figure*}

\begin{figure*}
\begin{lstlisting}[language=json]
{
  "CSP": "IBM",
  "data_type": "logs",
  "error": null,
  "governance_data": {
    "log_1": "Application started successfully.",
    "log_2": "Collecting system metrics.",
    "log_3": "Metrics collected successfully.",
    "log_4": "Sending data to API Gateway.",
    "log_5": "Data sent successfully.",
    "log_6": "Error handling and logging mechanism operational.",
    "log_7": "System monitoring and logging active.",
    "log_8": "Routine check completed successfully.",
    "log_9": "No errors detected in the last cycle.",
    "log_10": "All systems functional."
  },
  "log_id": "090f53c3-11f4-4871-a716-7357e6153259",
  "service_name": "cna-app",
  "timestamps": {
    "cna_timestamp": "2024-08-12T17:16:03.620091+00:00",
    "RG_GOV_IMS_API_Gateway_timestamp": "2024-08-12T17:16:03.689018+00:00",
    "RG_GOV_IMS_Converter_timestamp": "2024-08-12T17:16:03.770521+00:00",
    "RG_GOV_IMS_Archiver_timestamp": "2024-08-12T17:16:03.881582+00:00"
  }
}
\end{lstlisting}
\caption{Sample IBM JSON with data type logs}
\label{lst:json-ibm-logs}
\end{figure*}

\end{document}